\documentclass[conference]{IEEEtran}
\IEEEoverridecommandlockouts
\usepackage{cite}
\usepackage{amsmath,amssymb,amsfonts}
\usepackage{amsthm}
\usepackage{graphicx}
\usepackage{textcomp}
\usepackage{textcase}
\usepackage{wrapfig}
\usepackage{graphicx}
\usepackage{xcolor}
\usepackage{url}
\usepackage{tabularx}

\usepackage[linesnumbered,ruled,vlined]{algorithm2e}
\usepackage{algpseudocode}
\usepackage{listings}
\usepackage[symbol]{footmisc}

\lstdefinelanguage{cypherplus}{
  morekeywords={},
  keywordstyle=\color{red}\bfseries,
  classoffset=2,
  morekeywords={match,create,where,and,return},
  keywordstyle=\color{blue}\bfseries,
  classoffset=1,
  morekeywords={},
  keywordstyle=\color{Cerulean}\bfseries,
  classoffset=0,
  identifierstyle=\color{black},
  sensitive=false,
  comment=[l]{--},
  morecomment=[s]{/*}{*/},
  stringstyle=\color{purple}\ttfamily,
  morestring=[b]',
  morestring=[b]",
  commentstyle=\color[HTML]{216229}\textit,
  xleftmargin=1em
}
\newtheorem{definition}{Definition}[section]
\newtheorem{example}{Example}[section]

\makeatother
\newcommand{\tabincell}[2]{\begin{tabular}{@{}#1@{}}#2\end{tabular}}

\def\BibTeX{{\rm B\kern-.05em{\sc i\kern-.025em b}\kern-.08em
    T\kern-.1667em\lower.7ex\hbox{E}\kern-.125emX}}
\begin{document}

\title{PandaDB: Understanding Unstructured Data in Graph Database\\
\thanks{This work was supported by the National Key R\&D Program of China(Grant No. 2021YFF0704200). We also thank [...] for contributing [...].\\
$^*$Shen Zhihong and Tang Mingjie are the corresponding authors of this paper.\\
Shen Zhihong and Zhao Zihao contribute equally to this paper.
}
}

\author{\IEEEauthorblockN{Zhihong Shen$^{1,*}$, Zihao Zhao$^{1,2}$, Mingjie Tang$^{3,*}$, Chuan Hu$^{1,2}$, Huajin Wang$^{^1}$, Yuanchun Zhou$^{1,2}$}
\IEEEauthorblockA{\textit{Computer Network Infomation Center, Chinese Academy of Sciences$^1$}, Beijing, China \\
\textit{University of Chinese Academy of Sciences$^2$}, Beijing, China\\
\textit{Sichuan University$^3$}, Chengdu, China \\
bluejoe@cnic.cn, zhaozihao17@mails.ucas.ac.cn, mj.tang@scu.edu.cn,\\ huchuan19@mails.ucas.ac.cn, \{wanghj, zyc\}@cnic.cn}
}

\maketitle

\begin{abstract}
Unstructured data (e.g., images, videos, PDF files, etc.) contain semantic information, for example, the facial feature of a person and the plate number of a vehicle. There could be semantic relationships among data items. For example, a person's face may appear in two irrelevant photos. Also, part of data is in structured format (e.g., person's name and age). Naturally, end-users prefer to query unstructured data and structured data together based on the potential relationships among them. 
In this work, we build an open-source graph database named PandaDB to manage and query structured and unstructured data in graph. We first introduce graph as the data model to manage structured and unstructured data in one framework, then propose a query language extension to understand the semantic information of the unstructured data in the graph. Next, we develop a new cost model and related query optimization techniques to speed up the unstructured data processing in graph. Finally, we optimize the unstructured data storage and provide the index to speed up the query processing for unstructured data. PandaDB is widely used in industrial applications like FinTech, Knowledge Graph, and Recommendation System. The results show PandaDB can support a large scale of unstructured data query processing in a graph.
\end{abstract}


\section{Introduction}
\label{sec:introduction}

Both structured (e.g., numbers, strings) and unstructured (e.g., images, videos, PDF files, etc.) data describe the attributes of objects in various application domains, e.g., social networks, road networks, biological networks, and communication networks~\cite{gantz2011extracting}. 
The data of these applications can be viewed as graphs, where the nodes (a.k.a vertexes) and the relationships (a.k.a edges) have properties (a.k.a. attributes)~\cite{walid_sigmod,libkin2016querying}.
End users would prefer to issue queries for the graphs' topology, as well as the structured and unstructured data associated with the nodes and the relationships together.

\begin{figure*}
    \centering
    \vspace{-1.5em}
    \scalebox{0.9}{\includegraphics[width=2\columnwidth]{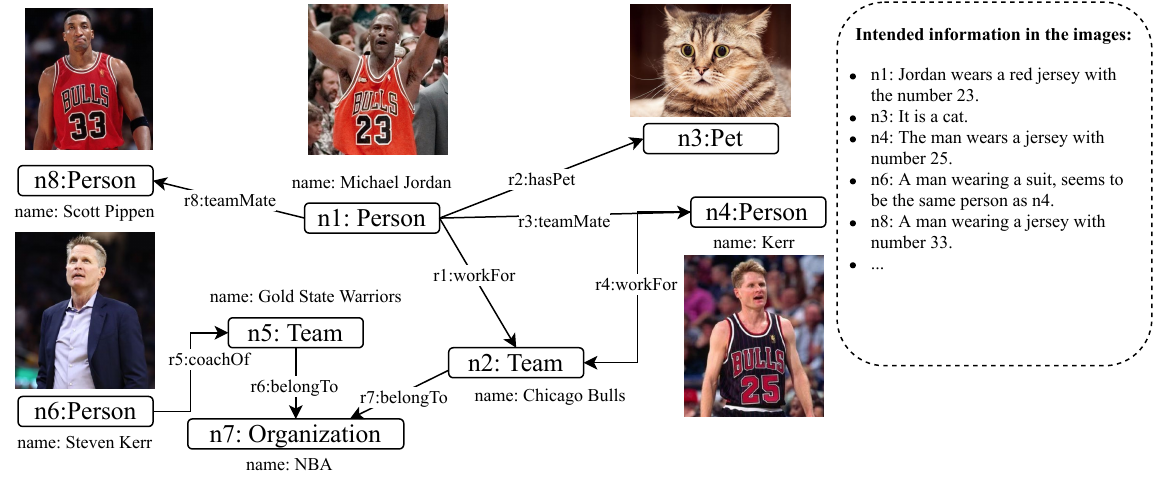}}
    \vspace{-1.0em}
    \caption{Example of querying unstructured data on graph}
    \vspace{-1.0em}
    \label{fig:example_graph}
\end{figure*}

Take Figure \ref{fig:example_graph} as an example, individuals (e.g., Michael Jordan) and related context information (e.g., NBA Chicago Bulls)  are represented as nodes in this graph. Then, the relationships between individuals (e.g., Michael Jordan works for Chicago Bulls) are viewed as the edges. In addition, the properties of nodes (e.g., $n_1$) in Figure~\ref{fig:example_graph} can be structured (birthday or name of Michael) or unstructured data (pictures, videos of Michael). End users usually initialize some queries to understand the data as 
follows:

\begin{example}
Graph data-related queries in Figure \ref{fig:example_graph}.
\begin{itemize}
\label{running_example_query}
\item $Q_1$: What is the color of Michael Jordan's pet cat?
\item $Q_2$: What jersey number did Michael Jordan's teammates wear at Bulls?
\item $Q_3$: Whether Kerr (Michael Jordan's former teammate) is the same person as the Gold State Warrior's coach Steven Kerr? 
\end{itemize}
\end{example}

To answer such queries (i.e., $Q_2$), traditionally, we at first find items with the name of \textit{Michael Jordan} from the database.
Then, we get \textit{Michael's} teammates at Bulls via the \textit{teamMate} relationship in the database.
Next, we fetch the corresponding teammates' photos from the file system and get the jersey numbers based on image information extraction models.
Finally, we return the basketball jersey numbers of \textit{Michael Jordan}'s teammates. 
As a result, developers often have to comprise multiple systems and runtime together. This gives rise to some issues such as managing the complexities of data representation, resource scheduling, and performance tuning across multiple systems.

In addition, we are facing several scenarios related to graph and unstructured data query processing in graph databases, as listed below. 

\begin{enumerate}
\item \textbf{Graph Neural Networks}:    
The Graph Neural Network (GNN) takes as input a graph endowed with node and edge features, then computes a aggregation results that depends on the features and the graph structure. Thus, the GNN builds representations of nodes and edges in graph data through neighborhood aggregation, where each node gathers features from its neighbors to update its representation of the local graph structure around it. Therefore, the GNN model needs to read the neighborhood properties of nodes, and related unstructured data information extraction (e.g., semantic understanding of an image, keywords extraction from a document, and voice messages) is useful to generate the features of GNN model training or prediction~\cite{michaelbeyond2022}. From the view of the graph database system, we need to propose a new data processor/operator to understand the semantic information of unstructured data in a Graph with hopes of accelerating graph deep learning models. 

\item \textbf{Fraudulent cash-out detection}: 
Credit card cash-out is attractive for investments or business operations, which are considered unlawful if exceeding a certain amount. Specifically, some credit card holders want to obtain cash through transactions, and the merchant receives the funds after transaction settlement by the acquires, then pays the funds back to the credit cardholder, charging the handling fee~\cite{LiuCYZLS18}. For one of our anonymous customers, we take the transaction among users as an edge, and each user as a vertex in the graph. Then, each transaction-related user signature is stored. Thus, we hope to identify the possible cash-out groups from the built graph and related users' signatures. Therefore, we want to discover the densely connected subgraphs (k-cores groups), and each pair of nodes of the subgraphs share similar human signatures.   
\end{enumerate}

To meet such demands in the real applications, we are facing challenges as listed below:
\begin{enumerate}
    \item This system needs to extend the data model in current graph databases and provide corresponding query semantic to manage unstructured data in graph databases. 
    \item The query processing engine should optimize the query plan based on the physical location of unstructured data and estimating the processing cost of unstructured data operators in graph databases. 
    \item The data storage in traditional graph databases should be extended to store unstructured data efficiently and provide a way to understand the meta data of unstructured data easily.
\end{enumerate}

The major contributions of this work are listed below:
\begin{enumerate}
    \renewcommand{\labelenumi}{\theenumi.}
    \item \textbf{Data model and query semantic}: We define the semantic and query operators for querying the content of unstructured data in a graph. This facilitates the graph query language to meet the description and query requirements of unstructured data without significant syntax changes. 
    
    \item \textbf{Query optimization}: We construct a cost model
    to formalize the query processing related to unstructured data processing in the graph and develop an optimizing algorithm to optimize the query plan.
    
    \item \textbf{Optimized data storage and indexing}: We optimize the physical storage of graph databases for supporting unstructured data management and develop a new type of index to speed up the queries for the unstructured data.
    
    \item \textbf{A new type of graph database system}: Based on the design mentioned above, a distributed graph database system, PandaDB, is implemented and tested for large-scale data.
\end{enumerate}

The remainder of this paper is organized as follows.
Section~\ref{sec:related_work} presents the related work. 
Section~\ref{sec:data-model-semantics} formalizes the data model and gives the query language.
Section~\ref{sec:overview} provides the system framework of PandaDB.
Section~\ref{sec:optimization} discusses the optimization of unstructured data queries.
Section~\ref{sec:datastorage} gives more details about the data indexing and storage.
The experiment results are presented in Section~\ref{sec:experiment}, and the conclusion is presented in Section~\ref{sec:conclusion}.
\section{Related work}
\label{sec:related_work}

\textbf{Graph database and processing systems}~\cite{sahu2017ubiquity,sahu2020ubiquity} have developed rapidly, flourishing in graph query and large-scale graph data management~\cite{wood2012query,walid_acm_graph,angles2008survey}.
For example, Neo4j~\cite{neo4j} and JanusGraph~\cite{janusgraph} are widely adopted for cloud and on-premise usage and focus on the querying and management of graph data~\cite{angles2018g,castellana2015memory,angles2017foundations}. 

Unlike structured data, users want to know the semantic information of unstructured data (e.g., text, photo, or video).
For example, regarding the plate number in the photo of a vehicle, the vehicle administration needs to find all cars with plate numbers starting with 123xxx.
To the best of our knowledge, the primary commercial products do not support the querying of unstructured data in big graphs~\cite{hugegraph,neo4j,janusgraph,martinez2011dex}.
In contrast to many existing systems that deal with batch-oriented iterative graph processing (i.e., graph analysis), such
as Pregel~\cite{pregel}, PowerGraph~\cite{gemini}, GraphX~\cite{graphx}, and Gemini~\cite{gemini}, PandaDB preserves the well-formed data model of the existing graph database research, and the extended declarative language allows user to understand the semantic of unstructured data. 

\textbf{Multimedia retrieval systems} support the querying of the content of unstructured data. However, most of the works are usually designed for a single data type and a specific retrieval purpose~\cite{chechik2008large,chua1995video,snoek2007adding,rui1999image,gudivada1995content}, such as face recognition~\cite{wagner2011toward,chen2000new} or audio speech recognition~\cite{shrawankar2013techniques}. 
Most of multimedia retrieval systems aim to improve the computer's ability to understand multimedia data, such as identifying more types of data and improving model accuracy. In addition, multimedia retrieval systems usually query unstructured data by tags or vector matching. As far as we know, most of multimedia retrieval systems do not query images or texts in one graph database framework. The related query plan of multimedia system can not be improved since it is lack of query plan optimization in database system.

\textbf{Collaborative retrieval systems} are usually built on the tools-chain-based system to support collaborative queries on structured data and unstructured data~\cite{wei2020analyticdb}.
A collaborative query is decomposed into several sub-queries on different modules.
Usually, a vector search engine is built for vector similarity search~\cite{johnson2019billion,douze2018link,zhang2019grip}, and a database system is prepared for structured data management.
In addition, an unstructured data analysis service is required to extract the feature vectors.
Then a data pipeline is built to connect these components.
Once the pipeline is implemented, the execution order of the components is fixed.
Thus, a pipeline could not support the flexible queries as the query pattern changes.
Data and related computation are distributed in different systems, and the consistency and correctness will take many resources to be maintained.
More importantly, the decoupled system framework loses the opportunity to optimize the workflow from beginning to end.
Some work~\cite{expensive_udf} trying to optimize the expensive UDFs in dataflows. However, this framework do not apply for a graph database system.

Currently, when users want to process unstructured data in a graph, they have to build a pipeline based on graph database system and different unstructured data processing models. As a result, users need to maintain the data consistency and query correctness in such a pipeline. More importantly, users need to tune such pipeline manually one by one, then the performance of pipeline is not guarantee in general. 
Based on such observation, we need to extend the current graph database to manage unstructured data in one system. So end users do not need to spend extra effort on how to process data rather than the query results itself. Then, the built system can optimize graph queries related to unstructured data efficiently and automatically.  
\section{Data model and Semantics}
\label{sec:data-model-semantics}
In this section, we first introduce the property graph and its query language, then introduce how PandaDB are extended to support the unstructured data querying in the property graph.

\subsection{Property Graph Model}
In the graph database community, data are typically represented as a property graph~\cite{angles2008survey,sahu2017ubiquity,sahu2020ubiquity}.
Every entity is represented as a node (a.k.a. vertex), identified by a unique identifier, having label(s) indicating its type or role.
The attributes of the entity are called properties of the node.
The relationship (a.k.a. edge) describes the association between entities. 
The nodes are connected by relationships.
A relationship starts at a node (namely source node) and ends at a node (namely target node). 
The category of the entity is taken as the node's label. A node could have more than one label.
We give the formal specification of the property graph data model as ~\cite{angles2008survey}. Details about the property graph model are shown in the tech report of PandaDB~\cite{tech_report_pandadb}.

\subsection{Graph Querying Language}
Cypher~\cite{francis2018cypher} is a standard graph query language that allows high-level and declarative programming for various graph operations, including graph traversal, pattern matching, and sampling. Due to page constraints, we provide an example of creating and querying data via Cypher for Figure~\ref{fig:example_graph} in the appendix of the tech report of PandaDB~\cite{tech_report_pandadb}.

The rich set of operators provided by Cypher makes it easy to
express a wide variety of graph computations.
However, the requirements of querying semantic information of unstructured data of graph nodes are still not met.

\subsection{PandaDB Extension}
\subsubsection{Unstructured Content Representation}
The properties of nodes/relationships in a graph can be unstructured and structured data.
In this work, we majorly focus on how to improve the query processing for unstructured data.
At first, we deem the semantic information of data as the \textbf{\textit{sub-property}}.
For example, in terms of node $n_1$ in Figure \ref{fig:example_graph}, the name and photo are the properties of $n_1$.
The printed number of the jersey is a sub-property of the photo. Obviously, an unstructured data item can have various potential sub-properties. For example, the jersey number and human facial features (e.g., color, hair, and eyebrow) in $n_1.photo$ are regarded as different sub-properties of Node $n_1$.
We formalize the sub-property definition as follows:
\begin{definition}
\label{sub-property-definition}
\textbf{Sub-property} is the semantic information in unstructured data, that is \\
\texttt{<data\ item> $\rightarrow$ subProperty = <semantic \ information>}
\end{definition}
\begin{example}
The semantic information of $n_1$'s photo in Figure~\ref{fig:example_graph} in represented as the following:
\begin{itemize}
    \item $n1.photo \rightarrow jerseyNumber = 23$
    \item $n1.photo \rightarrow face = <\$feature-vector>$
\end{itemize}
\end{example}
The list of sub-properties is pre-defined by the users, and it could be extended.

\subsubsection{Sub-property Acquisition and Filtering}
For the acquisition of semantic information of unstructured property, we introduce the sub-property extraction function $\phi$ :
\begin{definition}
\label{sem-info-definition}
Sub-property extraction function $\phi$: A finite partial function that maps a sub-property key to a sub-property value (semantic information) as following:
\begin{equation}
\begin{split}
&\phi: (N \cup R) \times \mathcal{K} \times \mathcal{SK} \rightarrow \mathcal{SV}, Sem \subset \mathcal{SV}\\
&\forall sv \in Sem, \exists ud \in S_{ud} \  where\  sv=\phi(ud, sk) 
\end{split}
\end{equation}
\end{definition}
Consider the nodes in Figure~\ref{fig:example_graph}, the name and the photo are the properties, and the \textit{face}, \textit{jerseyNumber} and \textit{animal} are the sub-property keys. The sub-property extraction in Figure~\ref{fig:example_graph}\ could be expressed as following ways:
\begin{itemize}
    \item 
    $\phi$($n_{1}$, photo, jerseyNumber) = 23\\
    $\phi$($n_{1}$, photo, face) = $<\$feature\_vector>$\\
    ...

\end{itemize}

Overall, a property graph including unstructured data is a tuple $UG = <G, \mathcal{SK}, \phi>$ where:
\begin{itemize}
    \item G is a property graph, whose property could be unstructured data.
    \item $\mathcal{SK}$ is a finite set, whose elements are referred to as the sub-property key of $UG$.
    \item $\phi$ is a function set, items of it are used to extract sub-property values from unstructured data.
\end{itemize}

\subsubsection{Query Language}

To query unstructured data in the property graph, we develop \textbf{\emph{CypherPlus}} to include new functions: \textbf{\textit{Literal Function}}, \textbf{\textit{Sub-property Extractor}}, and \textbf{\textit{Logical Comparison Symbols}}.

\textbf{\textit{Literal Functions}} treat the unstructured property as a BLOB (short for Binary Large Object) and create the unstructured property in a graph from a specific source. For example, \textit{Blob.fromURL(), Blob.fromFile()} and \textit{Blob.fromBytes()}, these functions are supplied by PandaDB. 

\textbf{\textit{Sub-property Extractor}} (represented as -\textgreater) is the semantic symbol of sub-property extraction function. It obtains the specific sub-property value from the data item, by invoking the sub-property extraction function.
The users define how to extract a specific sub-property from unstructured data, by binding the sup-property name with an AI model.

\begin{table}
\vspace{-1.5em}
  \begin{center}
    \vspace{-1em}
    \caption{Logical comparison symbols of unstructured data}
    \vspace{-1em}
    \scalebox{0.8}{
    \begin{tabular}{|c|c|c|}
    \hline
      \textbf{Symbol} & \textbf{Description} & \textbf{Example}\\
      \hline
      :: & The similarity between x and y. & x::y = 0.7\\
      $\sim:$ & Is x similar to y.  & x$\sim:$y = true\\
      $!:$ & Is x not similar to y. & x$!:$y = false\\
      $<:$ & Is x contained in y. & x$<:$y = true \\
      $>:$ & Is y contained in x. & x$>:$y = false \\
      \hline
    \end{tabular}
    }
    \label{table:logic-symbol}
  \end{center}
  \vspace{-2em}
\end{table}

\textbf{\textit{Logical Comparison Symbol}} offers a series of symbols as Table \ref{table:logic-symbol} to support logical comparison between sub-properties. 
According to predefined rules, these symbols are considered to compare logical relationships between specified semantic information. For example, when \textit{::} is used to compare face information, the similarity of two facial feature vectors is calculated. According to the definition of UDF in \cite{silberschatz2002database}, PandaDB allows users to define the semantic of operator (e.g. how to compute the similarity between two texts) in one framework, and provide one way to add new semantic symbols. Notice these symbols are read-only functions, data integrity would not be effected when invoking them.

Example~\ref{cypherplus-example} shows how the sub-property extractor and comparison symbol used in CypherPlus.
\begin{example}
\label{cypherplus-example}
We give the three graph queries for Figure \ref{fig:example_graph} as follows.
Note that the extensions of CypherPlus are in red.
\begin{lstlisting}[language=cypherplus]
-- Q1:What are the jersey numbers of 
-- Michael Jordan's teammates?
MATCH (n:Person)-[:teamMate]->(m:Person)
WHERE n.name=`{\color{black}'Michael Jordan'}`
RETURN m.photo`{\color{red}->}`jerseyNumber;
-- Q2:Is Michael Jordan's pet a Cat?
MATCH (n:Person)-[:hasPet]->(m:Pet)
WHERE n.name=`{\color{black}'Michael Jordan'}`
RETURN n3.photo`{\color{red}->}`animal = `{\color{black}'cat'}`;
-- Q3:Whether Michael Jordan's teammate Kerr the same person as 
-- Gold State Warrior's coach Steven Kerr?
MATCH (n1:Person)-[:teamMate]->(n4:Person), (n7:Person)-[:coachOf]->(n6:Team)
WHERE n1.name = `{\color{black}'Michael Jordan'}`
AND n4.name = `{\color{black}'Kerr'}`
AND n6.name = `{\color{black}'Gold State Warriors'}`
AND n7.name = `{\color{black}'Steven Kerr'}`
RETURN n4.photo->face `{\color{red}$\sim:$}` n7.photo->face;
\end{lstlisting}
\end{example}

The extraction, calculation and filtering of sub-property are different from the traditional UDF functions. Traditional UDF is a process of calculating and processing the data in the database without changing the existing data structure. 
The extraction, calculation and filtering of sub-property, however, are based on the semantic information of unstructured data objects. Its calculation rules focus on the semantic information of unstructured data objects, such as calculating the similarity of two faces. Semantic information itself is not data. It is only the information contained in the data.

\section{System Overview}
\label{sec:overview}
\begin{figure}
    \centering
    \vspace{-2em}
    \includegraphics[width=\linewidth]{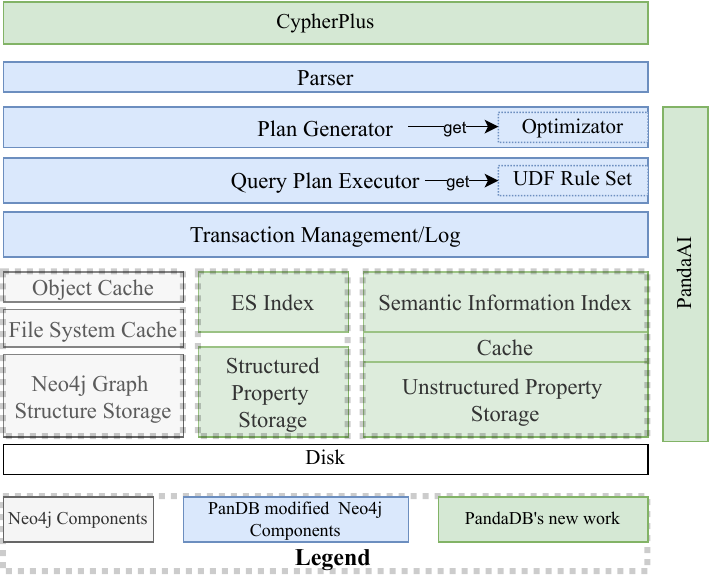}
    \vspace{-2em}
    \caption{Architecture of PandaDB}
    \vspace{-1em}
    \label{fig:architecture}
\end{figure}

As depicted in Figure~\ref{fig:architecture}, the PandaDB system adopts the native graph database Neo4j as the foundation.
Then the semantic-aware query parser, execution engine, and optimization algorithm are introduced, followed by the data storage and index to support efficiently querying structured and unstructured data.
Finally, an AI server is proposed for the execution runtime to understand the semantic information of unstructured data.

\subsection{Query Plan Optimization}
We modify the parser of Cypher to understand and parse the semantics of \textit{CypherPlus}. The newly added symbols are shown in Table~\ref{table:logic-symbol} besides the sub-property symbol $-\textgreater$.
In general, the execution plan of PandaDB is executed linearly one by one, following a conventional model outlined by the Volcano Optimizer Generator~\cite{graefe1993volcano}.
The query plan optimization applies standard rule-based optimizations, including constant folding, predicate pushdown, projection pruning, and other rules.
For example, to support querying the  properties of graph nodes, predicates of the property filtering operations are pushed down to the storage layer~\cite{levy1994query}.
This makes full use of the index in the storage layer.

As we introduced in previous section, understanding unstructured data always involves AI model inference, and this is time-consuming in real applications.
PandaDB estimates the cost of unstructured data operations in real time based on the cost estimation in in Section~\ref{sec:optimization}, and the optimizing algorithm is developed to optimize the corresponding query plan.

\subsection{Execution Operator}
\begin{table*}
    \centering
    \caption{Details about the unstructured data operators}
    \scalebox{0.9}{
    \begin{tabular}{|c|c|l|}
        \hline
        Operator & Arguments & Description \\
        \hline
        createFromSource() & URL or file path or binary content & Create a BLOB from the source. \\
        extract() & BLOB item \& sub-property name & Extract the sub-property(semantic information) from unstructured item. \\
        compareAsSet() & Two sets of semantic information & Compare the similarity of the semantic information in the sets. \\
        \hline
    \end{tabular}
    }
    \label{table:operators}
\end{table*}

A query is decomposed into different operators, and these operators are combined into a tree-like structure called execution plan. In this work, we introduce a series of new operators in Table~\ref{table:operators} to create unstructured data items from the data source, extract the sub-property and execute the logic computation.

In addition, we provide the user define function (UDF) for end-users to specify their way of understand the semantics of unstructured data. For example, users define a sub-property named \textit{face}. This represents the facial features of the individual photo. Next, our system can ingest the UDF (e.g., a face recognition model) to extract the facial features from the corresponding photos. 
We present a general interactive protocol (namely \textit{PandaAI}) between database kernel and AI models. Once a query obtains the semantic information from the AI model, the query engine sends a PandaAI-request to get the extracted information.
The server receives the request and extracts the semantic information using the model corresponding to the service asynchronously.
When the database query engine receives the extracted information, it caches the result and returns it to the user.

\subsection{Unstructured Data Storage In a Graph}
Graph storage is classified as non-native and native graph storage in the database community. For the non-native store, the graph storage comes from an outside source, such as a relational or NoSQL database. These databases store nodes and relationships of the graph without considering the topological, which may end up far apart in actual storage.

In this work, we opt for the native graph storage\footnote{https://neo4j.com/developer/kb/understanding-data-on-disk/}. The data is kept in store files for the native graph engine. Each file contains data for a specific part of the graph, such as nodes, relationships, node-related labels, and properties. Traditionally, the binary contents are stored as ByteArray in this storage. But this will degrades the IO performance greatly.
In the native graph storage, we introduce a new datatype named BLOB (binary large object) to store the unstructured data. It split the metadata and binary contents of unstructured data items, are load the unstructured data lazily.
More details about the unstructured data storage in graph are presented in Section~\ref{sec:datastorage}.

\subsection{PandaAI}
We use AI models to extract semantic information of unstructured data (namely sub-properties). At first, different semantic extraction functions have different input (e.g. images, videos, audios and texts) and output (e.g. numbers, strings, vectors). Thus, they are lack of unified abstraction at the system level.
Secondly, from the implementation level, the dependencies package of corresponding extraction functions are quite different. For example, Some functions may require special hardware with specific version (e.g. GPU version), thus, software dependency conflicts may appear in various plug-in functions.
Finally, the speed of an extraction function may vary greatly (e.g. cached or not, indexed or not). The wide scope of speed variation and the difference between  functions make it hard to optimize different models easily.
Therefore, PandaDB needs an AI service that can provide real-time extraction of semantic information, and the AI service should support flexible deployment to meet the changes in the demand of different resource. To meet this demand, we propose a component name as PandaAI to defines the interaction between AI services and database query engines.
\section{Query Plan Optimization} 
\label{sec:optimization}

This section first explains the procedure to generate the query plan for such a query, then formalizes a new cost model and algorithm to improve the query execution performance. The theoretical basis to optimize a query involving unstructured data is introduced in the appendix of the tech-report of PandaDB~\cite{tech_report_pandadb}.

\subsection{Query Plan Generation}
As introduced before, the design of \textit{CypherPlus} is motivated by Cypher~\cite{francis2018cypher}, XPath~\cite{kay2004xpath} and SPARQL~\cite{harris2013sparql}.
Given a query statement, the plan-generator generates the query plan based on the following steps: 
(a) Parses the query statement into an AST (Abstract Syntax Tree), checks the semantics, collects together different path matches and predicates.
(b) Builds a query graph representation of the query statement.
(c) Deals with the clauses and finds the optimal operator order. 
(d) Translates the optimal plan into the physical operators for data access.
Therefore, a query is decomposed into a series of operators, each of which implements a specific piece of work. 

In general, the query planning in PandaDB is optimized based on the IDP algorithm (an improved dynamic algorithm)~\cite{idp1,idp2} based on the corresponding cost model~\cite{gubichev2015query}. In this work, we extend this cost model and related algorithm to support unstructured data processing. 

These operators are combined into a tree-like structure (namely query plan tree, QPT).
Each operator in the execution plan is represented as a node in the QPT.
The execution starts at leaf nodes (usually \textit{AllNodeScan} or \textit{NodeScanByLabel}), and ends at the root node (usually \textit{Projection}).
The details of the basic query operator can be found in link
\footnote{https://neo4j.com/docs/cypher-manual/current/execution-plans/operator-summary}.
The query optimization in this work focuses on step(c) as mentioned above. It re-organizes the operators to find an optimal plan with less computation cost.
For an operator, its execution time depends on the data size of its input and its own characteristics.
\begin{figure}
    \centering
    \scalebox{0.85}{\includegraphics[width=\linewidth]{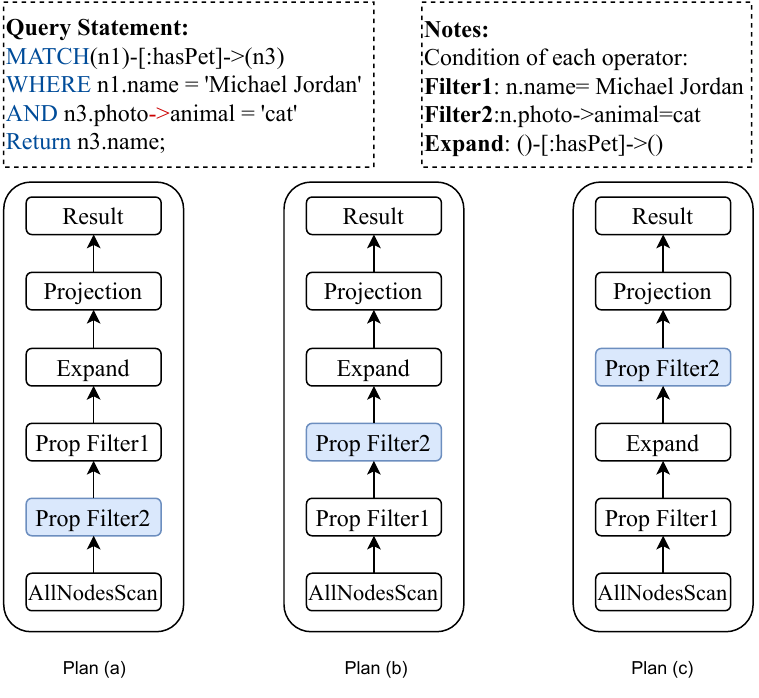}}
    \vspace{-1 em}
    \caption{Possible Execution plan of $Q_2$ in Example \ref{cypherplus-example}}
    \vspace{-1.5 em}
    \label{fig:execution}
\end{figure}

Consider the query statement in Figure \ref{fig:execution}, it queries \textit{Michael Jordan}'s pet cat's name.
The parsed operators include a structured property filter (Prop Filter1, filtering the data by the condition 'Michael Jordan'), an unstructured property filter (Prop Filter2, making sure the pet is a cat), then an expanding operator to find relevant nodes by node's relationships.
Also, there are some necessarily related algebra operations like \textit{Projection}, which is to select the expected result domain.
Figure \ref{fig:execution} shows three possible query plans to get the same queried results.
The difference between the three plans lies in the relative order between the operators.
Plan (a) executes sub-property filter (Prop Filter2) first, and then structured data filter(Prop Filter1). Next, expand on the filtered results.
Plan (b) executes the structured data filter (Prop Filter2) first, by contrary. 
Plan (c) executes the sub-property filter (Prop Filter2) at last.
However, the query execution time would differ.

In Plan(a), the \textit{Prop Filter2} filters the photos of all nodes in the database.
In Plan(b), the \textit{Prop Filter2} filters the output of \textit{Prop Filter1}.
While in plan(c) it only filters the output of the \textit{Expand} operator.
When \textit{Prop Filter2} is much slower than other operators, plan(c) will have the shortest execution time than others.
Because the fewer data the Prop Filter2 filters, the less time the whole plan takes if the Prop Filter2 is slower than other operators.
While in real-world applications, we could not suppose an unstructured filter is so expensive with cache and index, the query plan would also be more complex. Our system needs to adaptively optimize the query plan to obtain a fast execution plan by considering processing unstructured data in the graph database. 

\subsection{Query Plan Optimization For Unstructured Data Querying}
\label{subsec:planoptimization}

Cost based plan optimization (short as CBO) at first estimates the cost of query operations based on a cost model (usually a cardinality model), then apply optimization algorithm to compute an newly generated plan.
In term of unstructured data query processing, we at first check whether there is a cache and index for semantic information. If the data is cached, we do not need to repeat the semantic information extraction process. On the other hand, online extraction of semantic information for unstructured data is performed. Note this is time consuming in general. As a result, the related query processing time would fluctuates greatly and this is affected by data caching, workload, data distribution and other factors.
So we need to extend the CBO of existing graph databases. PandaDB at first introduces a new method to calculate the expected speed of an unstructured data operator (e.g., property filter) on the fly, then optimizes the query plan with a greedy strategy as introduced below. 

At first, we observe that semantic information are cached/indexed or not would influence the performance greatly. When such semantic information are cached and indexed, unstructured data operator shows much better performance than the ordinary operations (e.g. \textit{NodeScan} and \textit{PropertyFilter}) in the graph databases, then the corresponding plan optimization need to adjust this cost update. Base on this observation, the CBO of PandaDB is computing the related runtime cost gradually and update the cost one by one. 

Based on the designs mentioned above, Definition~\ref{definition-cost-model} formalizes the cost model in this work as following. 
\begin{definition}
\label{definition-cost-model}
Given an unstructured data operator $p$ and its speed when it is called at the $i$th time, the computation cost of operator $p$ is estimated as following way:
$$\mathbb{E}(\varepsilon(\sigma{_p})) = \mathbb{E}(v_{i+1}(\sigma{_p})|v_i(\sigma{_p})) * \mathbb{E}(|T|)$$
\end{definition}
Where the $\mathbb{E}(|T|)$ is the expected size of the input table $T$. 
Given $v_i(\sigma{_p})$, the expected cost when $p$ is called at the ($i+1$)th time is calculated by the following way.
$$
\mathbb{E}(v_{i+1}(\sigma{_p})|v_i(\sigma{_p})) = \frac{v_{0}(\sigma{_p}) + k*v_{i}(\sigma{_p})}{k+1}
$$

Where $v_{0}(\sigma{_p})$ is the initial speed of $p$, and $k$ is the adjust factor based on following observation. When the semantic information of newly added data are not cached and indexed, 
it brings a bias between estimated result $v_i(\sigma{_p})$ and actual speed.  
So we introduce $k$ to adjust the cost estimation to the newly added data. In other words, the higher $k$ means that the previous recorded speed $v_i(\sigma{_p})$ contributes more to the most recent estimation $v_{i+1}(\sigma{_p})$. On other hand, the lower value of $k$ means the previous result has lower impact. 
$k$ is a hyperparameter and is defined via database admin, we take $k$ equal to five based on our experience in this work. 
Notice that when an unstructured data operator $p$ is never used before, we thought it was much times slower than structured data operation. Thus, initial estimation $v_0(\sigma_p)$ is equal to $\beta$ times of the average speed of structured data operators, where $\beta$ is computed based on historic data.  In the future, we hope to tune the value $k$ and $\beta$ based on online learning approaches. 

After computing the approximate cost of unstructured data processing, we adopt a greedy strategy to optimize the query plan based on the aforementioned cost estimation. The optimization is illustrated in Algorithm \ref{optimization-algo}.
It employs a  \textit{PlanTable}, which keeps the latest constructed logical plans in the recursion of the optimization, and the \textit{Cand} maintains the operators have not been added in the \textit{PlanTable}. An entry of \textit{PlanTable} contains a logical plan that covers a certain part of the query graph (identified by IDs of nodes in that subgraph), along with the cost of the plan and its cardinality.

At first, the proposed algorithm inserts all the leaf plans (node scan, join, projection or expand) into the \textit{PlanTable} (lines 3-5).
The leaf plans are constructed according to the query graph $\mathcal{Q}$, each node in $\mathcal{Q}$ is transferred into a leaf plan.
Besides, the essential join, projection, and expand operations are constructed as leaf plans.
So these leaf plans should cover all nodes in the query graph $\mathcal{Q}$. 
And then, it repeats the greedy algorithm (lines 6-8) until it gets a query that is complete and covers the whole query graph $\mathcal{Q}$.
The \textit{GreedyOrdering} collects the candidate solution formed by joining a pair of plans from the \textit{PlanTable} (lines 12-16) or expanding a single plan via one of the relationships in the query graph (lines 17-19).
Next, \textit{GreedyOrdering} picks up the best candidate plan, inserts it into $\mathcal{P}$, and deletes all the plans from P which are covered by the best plan (lines 22-24).
Note that the best candidate plan is the plan which has the min estimated cost.
The procedure is stopped as soon as there are no candidates to consider.
At this point, the \textit{PlanTable} will contain a single plan that covers all the nodes, which we return as a result.

\begin{algorithm}
    \SetKwFunction{Main}{OptimizationFunc}
    \SetKwFunction{Greedy}{GreedyOrdering}
    \SetKwFunction{Pick}{PickBest}
    \SetKwProg{Fn}{Function}{:}{}
    \caption{Cost-based Plan Optimization Algorithm}
    \label{optimization-algo}
    \KwIn{Query graph $\mathcal{Q}$, Statistic information $\mathcal{S}$}
    \KwOut{Query plan P that covers $\mathcal{Q}$}
    
    \Fn{\Main{$\mathcal{Q}$, $\mathcal{S}$}}{
        $\mathcal{P}$ $\gets$ $\emptyset$ \Comment{PlanTable}\\
        \For {n $\in$ $\mathcal{Q}$}{
            $T \gets leafPlan(n)$ \\
            $\mathcal{P}$.insert(T)\\
        }
        Cand $\gets$ \Greedy{$\mathcal{P}$,$\mathcal{S}$}\\
        \While{size(Cand) $\geqq$ 1}{Cand $\gets$ $\Greedy{a,b}$}
        \textbf{return} $\mathcal{P}$
    }
    
    \Fn{\Greedy{$\mathcal{P}$,$\mathcal{S}$}}{
        Cand $\gets \emptyset$ \Comment{Candidate Solutions}\\
        \ForEach{$P_1$ $\in$ $\mathcal{P}$}{
            \ForEach{$P_2$ $\in$ $\mathcal{P}$}{
                \If{CanJoin($P_1$, $P_2$)}{
                    T $\gets$ constructJoin($P_1$, $P_2$)\\
                    Cand.insert(T)
                }
            }
        }
        \ForEach{$P_1$ $\in$ $\mathcal{P}$}{
            T $\gets$ constructExpand($P_1$)\\
            Cand.insert(T)
        }
        \If{size(Cand) $\geqq$ 1}{
            $T_{best}$ $\gets$ pickBest(Cand, $\mathcal{S}$) \Comment{Pick the best plan}\\
            \ForEach{T $\in$ $\mathcal{P}$}{
                \If{covers($T_{best}$, T)}{$\mathcal{P}$.remove(T) \Comment{Delete covered plans}}
            }
            $T_{best}$ $\gets$ applySelections($T_{best}$)\\
            $\mathcal{P}$.insert($T_{best}$)
        }
        \textbf{return} Cand;
    }
    
    \Fn{\Pick{Cand, $\mathcal{S}$}}{
        C $\gets$ $\emptyset$ 
        \Comment{Record estimated cost of each table} \\
        \ForEach{T $\in$ Cand}{
            est $\gets$ cost($\mathcal{S}$, $\mathcal{T}$, $\mathcal{O}$)\\
            C.insert(est, T)
        }
        $T_{best}$ $\gets$ min(C)\\
        \textbf{return} $T_{best}$
    }
\end{algorithm}

\begin{figure}
    \centering
    \includegraphics[width=\linewidth]{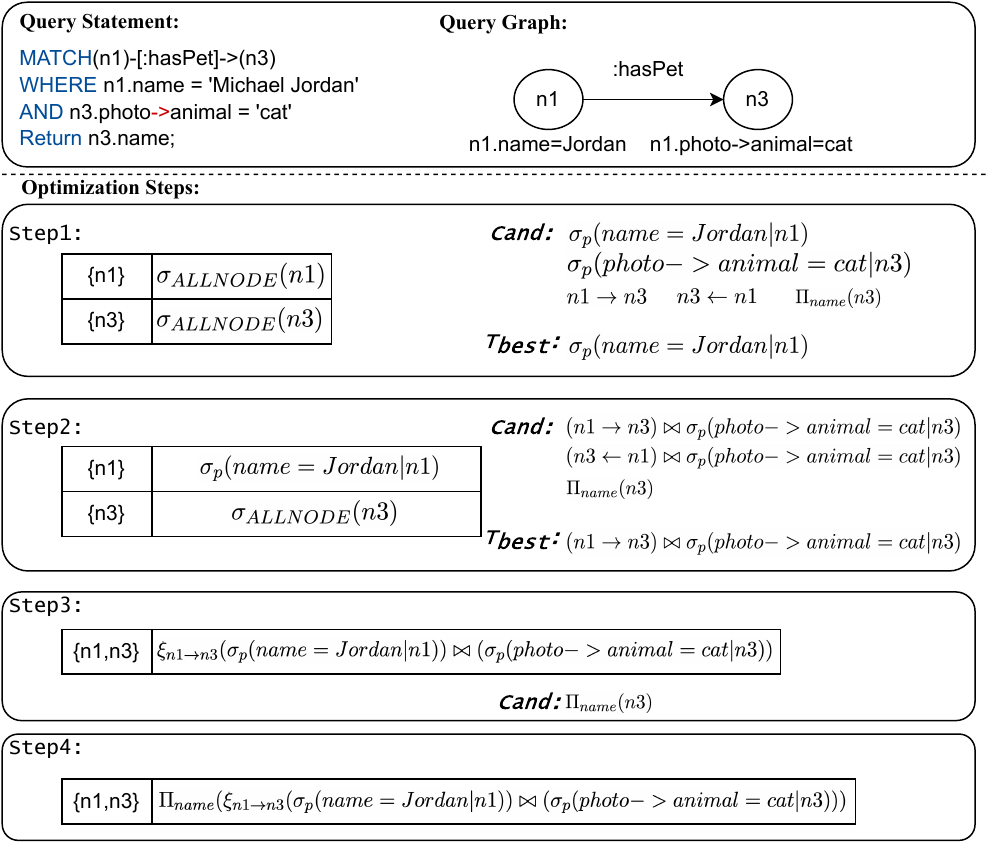}
    \vspace{-1 em}
    \caption{Example of Optimization Steps for Query2}
    \label{fig:opti-steps}
\end{figure}

\textbf{Running example}. 
To give an example for Algorithm \ref{optimization-algo}, Figure \ref{fig:opti-steps} gives a query statement and its query graph.
The figure shows the \textit{PlanTable}, Cand, and $T_{best}$ step-by-step.

Step1: The table is initialized with the plans that offer the fastest node access. 
This query does not specify the label of nodes, so the table could only obtain the nodes by plain \textit{AllNodeScan}.
The filter operations and projection are added into \textit{Cand}.
There are only two possible paths to expand: $n_1 \rightarrow n_3$ and $n_3 \leftarrow n_1$.
The former means to start from $n_1$, expand by the out-relationship, while the latter means to start from $n_3$, expand by the in-relationship.
They are added into the \textit{Cand}.
Supposed that the filter by name is the best candidate in \textit{Cand}, it is inserted into the \textit{PlanTable}.
This operation covers the \textit{AllNodeScan} of $n_1$, so the \textit{AllNodeScan} is removed.

Step2: The two expand operations could be joined with the filter operation.
Supposed the first in \textit{Cand} is the best candidate, insert it into the \textit{PlanTable}.
The ($n_1 \rightarrow n_3$) is represented as $\xi_{n_1 \rightarrow n_3}$.
The result covers the plain \textit{AllNodeScan} of n3, so it is removed from the \textit{PlanTable}.
Then goes to Step3; the only candidate left is the projection; insert it into the \textit{PlanTable}.
The final query plan is shown in the \textit{PlanTable} of Step4. It is the algebra representation of the query plan shown in Figure~\ref{fig:execution} (b).

\textbf{Complexity analysis}. The greedy procedure (lines 6-8) starts with n plans and removes at least one plan at every step. So it is repeated at most n times, where n is the count of nodes in the query graph $\mathcal{Q}$.
The complexity of estimating the cost of an unstructured property filter is $O(1)$.
Then, assuming that \textit{canJoin} utilizes the Union-Find data structure for disjoint sets, the complexity of the entire algorithm becomes $O(n^3)$.
In consideration about the value of n (not exceed 20, usually), this complexity is acceptable.

\section{Data storage and indexing}
\label{sec:datastorage}
\begin{figure}
    \centering
    \includegraphics[width=\linewidth]{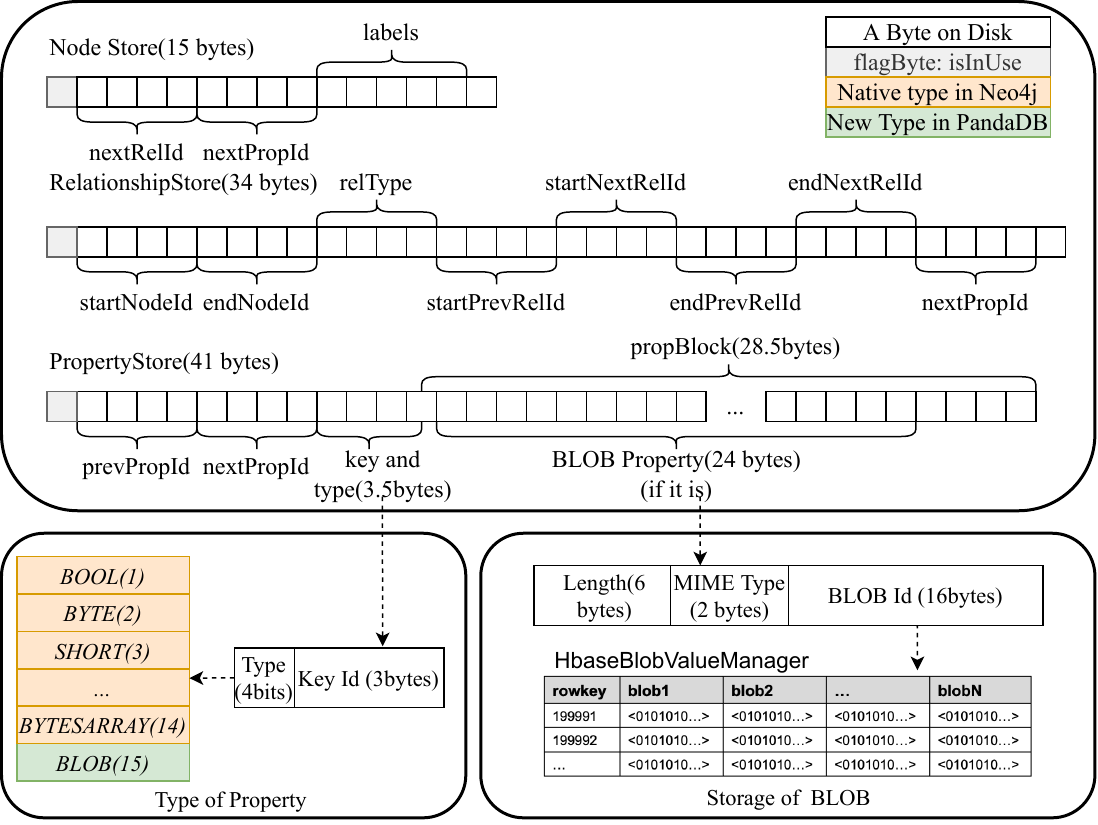}
    \caption{Data Storage on Disk}
    \vspace{-1.5 em}
    \label{fig:storage}
\end{figure}

In order to realize query optimization for graph and unstructured data, query engine comes to optimize the plan according to different kinds of statistic information, for example, data distribution and the cost to access data, etc. Currently, most of graph databases like Neo4j store the content of unstructured data in the form of \textit{ByteArray}, thus, it can not understand the metadata of unstructured data before deserializing the data.
In this section, we first introduce how to store the graph structure data and property data (including structured and unstructured data) in PandaDB. Then, we motivate the newly developed indexing to speed up the query processing for unstructured data in the graph database.

\subsection{Support Unstructured Data Storage In Graph Database}
PandaDB modifies the Neo4j storage\footnote{https://neo4j.com/developer/kb/understanding-data-on-disk/} to support unstructured data storage in a graph.
Figure~\ref{fig:storage} lists the related data storage format. \textit{Nodestore} uses the \textit{nextRelId} and \textit{nextPropId} to store the physical address of the relationship and property for the corresponding node.
Similarly, \textit{Relationshipstore} stores the address of \textit{startNodeId} and \textit{endNodeId}, where \textit{startNodeId} and \textit{endNodeId} are the related nodes of this relationship.

\textit{Properties} are stored as a double-linked list of property records, each holding a key and value and pointing to the next property. For example, \textit{propBlock} is used to store the content of the property in binary format. Originally, users stored the unstructured data in the block \textit{propBlock} as a \textit{ByteArray}. 
This storage shows success in real applications with the wide spread of Neo4j.
However, this way may not be a good solution for storing unstructured objects in graphs. 
First, unstructured data objects include metadata, such as MIME Type, length, version, etc. If the contents of the objects are stored in \textit{ByteArray} only, the related metadata will be lost. If these metadata are also encoded into \textit{ByteArray}, additional deserialization operations are required in each query execution, which will obviously decrease performance. Secondly, some queries often only need to read the metadata or the parts of bytes for the unstructured data object. Therefore, loading the whole \textit{ByteArray} into memory is unnecessary in most cases.

In this work, we modify the format of \textit{property} and introduce the binary large object (BLOB) as a new datatype to store the unstructured data.
From the bottom of Figure~\ref{fig:storage}, the metadata (i.e., length, MIME type, and id) of BLOB are stored in the last 28.5 bytes.
For those BLOBs under 10kB, the binary content is stored in another file, like a long string and array storage.
For those over 10kB, storing it into a native file will influence the performance because the BLOBs will be fully loaded into the memory.
So we adopt HBase to maintain the BLOBs.

Overall, PandaDB stores unstructured data in the following ways:
(1) Treat the unstructured property as a BLOB.
(2) Store the metadata and literal content of the BLOB, respectively.
(3) The metadata (including length, MIME type, and the id of BLOB) are kept in the property store file, as shown in Figure \ref{fig:storage}.
(4) For those BLOB whose literal value is less than 10kB, store it in the same method as long strings.
(5) For that exceeds 10kB, store them in the \textit{BLOBValueManager} based on HBase. The \textit{BLOBValueManager} organizes and manages BLOB in a BLOB-table, which has n columns.
In a row of the BLOB-table, each column stores a BLOB literal value.
The location of a BLOB could be calculated by its Id by the following formula, where $|column|$ means the count of the columns in HBase:
\begin{equation}
\begin{split}
& row\_key(BLOB) = id(BLOB)/|column|   \\
& column\_key(BLOB) = id(BLOB)\%|column|
\end{split}
\end{equation}
The \textit{BLOBValueManager} could quickly locate a BLOB by its id, as shown in Figure \ref{fig:storage}.
Besides, the transmission of BLOB between \textit{BLOBValueManager} and Query Engine is streaming.

\subsection{Semantic Information Cache and Indexing}
The semantic information of unstructured objects could be either lazily or eagerly computed and stored, perhaps indexed in the eager case.
The lazy case would not require a semantic information cache. Thus the system does not need to maintain the version of cached information and keep the consistency.
The eager case would require a caching mechanism to maintain the semantic information while do not need to compute the semantic information on the fly.
\begin{figure}
    \centering
    \includegraphics[width=\linewidth]{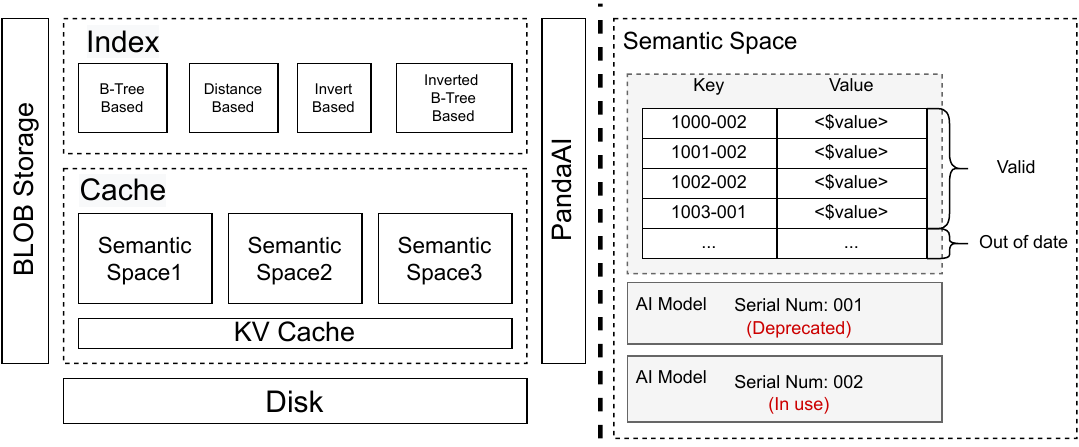}
    \caption{Cache and Index of Semantic Information}
    \vspace{-1.5 em}
    \label{fig:cache-and-index}
\end{figure}

\subsubsection{Semantic Information Extracting and Caching}
PandaDB extracts semantic information and stores it via the Key-Value format, where the key is composed of the id of the unstructured data item and the serial number of the AI model.
Moreover, the value is the semantic information. The system first tries to query the cache for each query including semantic information.

Figure \ref{fig:cache-and-index} shows the cache mechanism. Naturally, one AI model indicates one semantic space (one-to-one mapping). 
When the admin updates the AI model of a semantic space, the new model would have an updated serial number. A cache is valid when the serial number in the cache's key equals the latest model.
For example, suppose that the AI model with serial number 002 is in use, then the fourth cache is out of date.
Because the serial number of it is 001.

\subsubsection{Semantic Information Index}
Each kind of semantic information has its own meaning and format.
For example, facial features are vectors, text contents of audios are in the string format, etc.
In this work, PandaDB is able to choose a suitable index for corresponding semantic information automatically.
For the numerical data, the semantic index is based on B-Tree~\cite{btree1,btree2}. Then, an inverted index~\cite{inverted1,inverted2} is adopted for semantic information under the format of strings and texts.
For high dimensional vector data, we adopt inverted vectors search~\cite{invertedvectorsearch}. 
With the help of the index, the query plan generator is able to push down the related semantic information operator into the index. This greatly speeds up the data query processing. In addition, PandaDB applies two strategies for building indexes, batch building and dynamic building. The former applies to a semantic space that is not indexed before or after the corresponding AI model is updated.
The latter is adopted when there is a new semantic information item(i.e., a newly added unstructured data item in the database). More details are given in the appendix of the tech report of PandaDB~\cite{tech_report_pandadb}. 

\section{Experiment and implementation}
\label{sec:experiment}

\subsection{Implementation}
PandaDB extends Neo4j to support unstructured data processing. We modify the query parser, plan generation optimization and related data storage. More than 50k lines of scala code are added. All the source codes could be accessed at the link\footnote{https://github.com/grapheco/pandadb-v0.1}.
In addition, we choose \textit{HBase}~\cite{bigtable} to store unstructured data.
We implement the semantic information index engine adopting \textit{Milvus}~\cite{milvus}, an open-source C++-based vector database for vector search.
In addition, PandaDB adopts \textit{ElasticSearch}\cite{gormley2015elasticsearch,kononenko2014mining} as the index for structured property data. 
As introduced before, the property filter is pushed down to be executed on the \textit{ElasticSearch}. 

In order to improve throughput and availability, PandaDB replicates data to multiple nodes. When a new physical driver connects to a cluster, the queries it sends are divided into reading-query and writing-query. Thus, the reading-query only reads the data, while writing-query also modifies the data. Reading-query is randomly distributed to any available machine, and writing-query is forwarded to the leader for execution. The leader node initiates data synchronization within the cluster.

When the leader node executes a writing-query, it records its corresponding query statements and assigns a version number to each writing-query in ascending order. The version number and query statement are recorded in the log. This log is synchronized to other nodes in the cluster.
When a node goes online, it first compares whether the local log version is consistent with the log version of the leader in the current cluster. If consistent, the node can join the cluster. If the local log version is lower than the cluster log version, execute query statements in the local log until the version is consistent.

\subsection{Experimental Setup}
\label{subsec:expr-setup}
We evaluated PandaDB to verify the effectiveness of the proposed designs, as well as its performance improvement over existing native solutions.
We design eight typical query statements to simulate queries in real-world applications.
In terms of each query, we compare the execution time of the native solution, PandaDB without optimization, and PandaDB with optimization, by considering whether the semantic information is cached or not, respectively.
The performance improvement of PandaDB is mainly reflected in the query execution time, not the accuracy.
We design the experiment with respect to Ref~\cite{experiment_1} and Ref~\cite{experiment_2}.

\subsubsection{\textbf{Dataset}}
We combine a graph benchmark dataset and a face recognition dataset to obtain a property graph including unstructured data (images); 
both datasets are public to download.
For property graph data, we adopt the Linked Data Benchmark Council Social Network Benchmark (LDBC-SNB)~\cite{erling2015ldbc}, 
a scalable simulated social network dataset organized as a graph.
For unstructured data, we use Labeled Faces in the Wild (LFW)~\cite{LFWTech}, a public benchmark for face verification.
We attach the photos in LFW to person nodes in LDBC-SNB, each node a photo.
For recording the mapping between node and photo, the photo's id is set as a property of the node.
We use the different scales of datasets to evaluate the performance of PandaDB. The datasets are detailed in Table~\ref{table:dataset}, where SF is short for scale factor, an argument to describe the scale of datasets.

\subsubsection{\textbf{Testbed}} 
The experiments are conducted with a cluster including five physical machines.
Each node has 52 logical cores, 128GB RAM, 2TB SSD, and 215TB HDD. We also conduct the same experiment studies over the Alibaba Cloud and Amazon AWS cloud. The experiment results share a similar trend to the private cluster results. 

\begin{table}
    \label{dataset_table}
    \centering
    \caption{Details about the dataset}
    \scalebox{1}{
    \begin{tabular}{|l|c|c|c|c|}
        \hline
        Name & \#Node & \#Relationship & \#BLOB & Total Space\\
        \hline
        SF1& 3115527 & 17236389 & 9916 & 2.0GB \\
        SF3 & 8879918 & 50728269 & 24292 & 5.6GB \\
        SF10 & 28125740 & 166346601 & 65675 & 18.0GB\\
        SF30 & 83815107 & 510116996 & 165643 & 57.2GB\\
        SF100 & 266439718 & 1682136459 & 430671 & 164.9GB\\
        \hline
    \end{tabular}
    }
    \label{table:dataset}
    \vspace{-0.5em}
\end{table}

\subsubsection{\textbf{Query}}
\label{subsubsec:query}
The experiment modifies eight graph queries to include the unstructured data processing to test the system performance.
We only detail four of the eight queries as follow since the others share the same trend as queries in Section ~\ref{query_list_1}.
Note that the symbol $\sim:$ is defined to judge whether two faces are similar by comparing the similarity between the facial features.
\begin{lstlisting}[language=cypherplus]
-- Q1: Query a node by name and photo. 
Match (n:person) WHERE n.photo `{\color{red}$\sim$:}` Blob.fromURL('$url') AND n.firstName = '$name' RETURN n;
-- Q2: Query the shortest path between two nodes.
MATCH (n:person),(m:person) WHERE m.photo `{\color{red}$\sim$:}` Blob.fromURL('$url') AND n.firstName = '$name' RETURN shortestPath((n)-[*1..3]-(m));
-- Q3: Whether two nodes refer to the same person.
MATCH (n:person),(m:person) WHERE n.firstName='$name1' AND m.firstName='$name2' RETURN n.photo `{\color{red}$\sim$:}` m.photo;
-- Q4: Whether the two friends looks similar.
MATCH p = (n:Person)-[:friendOf]->(m:Person) WHERE n.photo `{\color{red}$\sim$:}` m.photo RETURN p;
\end{lstlisting}
\label{query_list_1}

\subsubsection{\textbf{Baseline}} 
We implemented four query processing based on queries listed in Section VII-C. Four workflows are detailed as below:
\begin{enumerate}
    \item Q1: Find the photos whose facial features are similar to those of the specific BLOB. Next, retrieve the corresponding nodes of the photos, then filter the nodes by the \textit{firstName}.
    \item Q2: Find the nodes whose \textit{photo} is similar to the specific BLOB and the nodes whose \textit{firstName} meets the argument. Then retrieve the shortest path between the nodes in the graph database.
    \item Q3: Retrieve the nodes whose \textit{firstName} meets the arguments in the query statement, then calculate the similarity of the facial features.
    \item Q4: Fetch the nodes corresponding to the path, then calculate the similarity of the facial features.
\end{enumerate}

To implement pipelines sharing the same functionality as graph queries in the previous section, we build a data processing pipelines based on Neo4j, PandaAI and a file system. For example, the workflow extracts the face vectors in the photos (if not cached), then calculates the similarity between faces according to the input, finally returns the result when the related similarity is bigger than the predefined threshold.
Naturally, when the similarity of two commercial features exceeds a predefined value, two faces are regarded as the same individual. More details about native solution for each queries are presented in tech-report of PandaDB~\cite{tech_report_pandadb}. 

\subsection{Data Import}
Data import is to load data into the graph via batch. The time-consuming of importing all data (including image data) between PandaDB and Neo4j is compared. Neo4j stores the picture data as ByteArray. The test results are shown in Table \ref{data_import}. PandaDB saves more than 20 percent overhead compared with Neo4j while importing unstructured data.

\begin{table}
    \centering
    \caption{Data Import Time}
    \vspace{-1em}
    \scalebox{0.8}{
    \begin{tabular}{|l|c|c|c|}
        \hline
        DataSet & \tabincell{c}{Baseline(s) \\ (No Images)} & PandaDB(s) & Neo4j(s) \\
        \hline
        SF1   & 30.8  & 32.6  & 33.8 \\
        SF3   & 62.6  & 61.3  & 69.5 \\
        SF10  & 131.5 & 138.1 & 154.6\\
        SF30  & 344.3 &  429.3& 541.1 \\
        SF100 & 1898  & 2030 & 2577 \\
        \hline
    \end{tabular}}
    \label{data_import}
    \vspace{-1.5em}
\end{table}

\subsection{System Latency and Throughput}
In order to test the latency and throughput of PandaDB, we use Apache JMeter\footnote{https://jmeter.apache.org/} to simulate concurrent requests in real applications. We adopt the default setup of JMeter in this work. In this experiment, we find that the average latency of a single query keeps around 20ms. This latency is acceptable in most industrial applications. At the same time, the QPS of PandaDB reaches 5300 under the current resource configuration. This is enough for most applications. In addition, the QPS of PandaDB can increase as more computation resources are added due to the nature of the high-available architecture of this system. 

\subsection{Effect of Query Optimization}
\begin{figure*}
    \centering
    \vspace{-2 em}
    \scalebox{0.85}{\includegraphics[width=2\columnwidth]{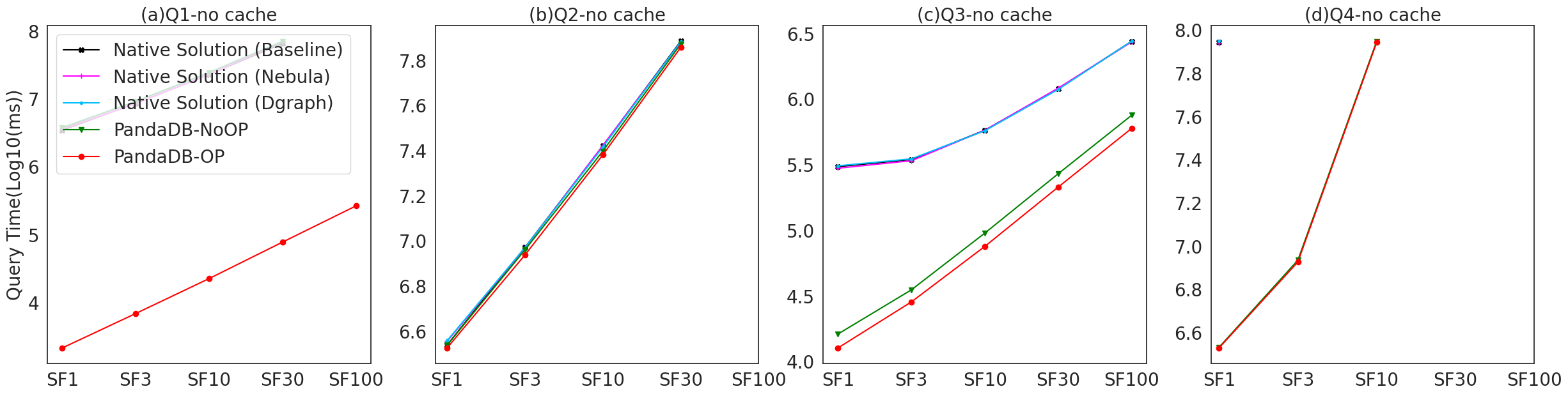}}
    \vspace{-1 em}
    \caption{Performance study of optimization without semantic information cache.}
    \vspace{-1em}
    \label{fig:expr_overview-nocache}
\end{figure*}

\begin{figure*}
    \centering
    \scalebox{0.85}{\includegraphics[width=2\columnwidth]{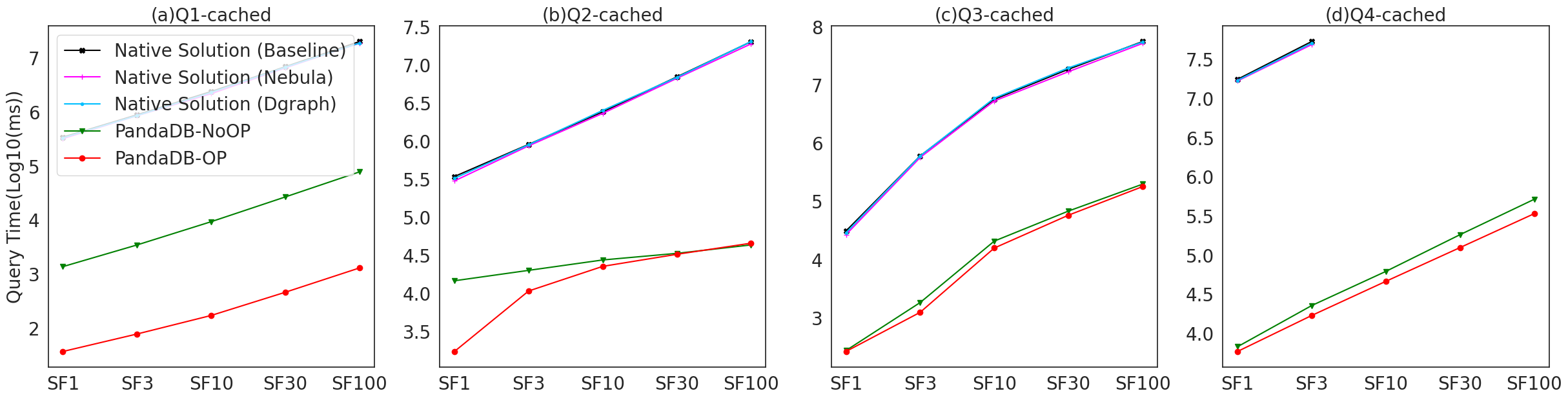}}
    \vspace{-1 em}
    \caption{Performance study of optimization with semantic information cache.}
    \vspace{-1em}
    \label{fig:expr_overview-cached}
\end{figure*}

We study the query performance based on queries in Table \ref{table:dataset}.
The results are shown in Figure \ref{fig:expr_overview-nocache} and Figure \ref{fig:expr_overview-cached}.
The x-axis means the scale of the dataset, the details about the scale are introduced in Section \ref{subsec:expr-setup}.
The y-axis means the execution time, the shorter, the better, and we take the logarithm of the execution time in the figures because of the significant performance gap.
The \textit{PandaDB-NoOP} stands for a PandaDB without optimization for unstructured data queries processing.
And the \textit{PandaDB-OP} is optimized for unstructured data queries by the method introduced in Section \ref{subsec:planoptimization}. We provide three native solutions for each queries and each native solutions share the same procedure. The only difference among three implementation is the graph databases system (e.g., Neo4j\footnote{https://github.com/neo4j/neo4j}, Nebula\footnote{https://github.com/vesoft-inc/nebula} and Dgraph\footnote{https://github.com/dgraph-io/dgraph}).

We set the upper limit of query time to 24 hours.
When the execution time of a query exceeds 24 hours, we regard the query times out and it will not show the result in the figure.
For example, the baseline times out on Q4 over all the datasets when the semantic information is not cached (i.e., Figure \ref{fig:expr_overview-nocache}(d)).

From the Figure~\ref{fig:expr_overview-nocache} and Figure~\ref{fig:expr_overview-cached}, we observe that performance of three graph databases based on native solution is dominate by (a) moving data among different components (b) cost of extracting semantic information from unstructured data. As we introduced before, if the semantic information is cached, the cost of moving data among different components should be the major bottleneck. On the other hand, the cost of extracting semantic information from unstructured data would be other major issue. Three native solutions do not have significant difference in the overall query processing time. On the other hand, PandaDB can process the data in one pipeline and optimize the query processing based on the cost model, then it wins 100x speedup against the native solution on average. 

When the semantic information is not cached, PandaDB has about three orders of magnitude advantages over the native solution for Q1;
PandaDB is 10x faster than the baseline on average for Q3.
For Q4, the native solutions failed to finish the query on all four datasets within the time limitation.
Compared with Q1, Q3 and Q4, PandaDB has less performance advantage in Q2.
The query optimization allows PandaDB to execute the query with fewer extraction operations.
Actually, according to the optimization detailed in Section \ref{sec:optimization}, PandaDB filters the data according to the structured data and then filters the result by semantic information.
But the native solutions have to filter all the semantic information.
While in Q2, both PandaDB and the native solutions need to extract semantic information of all the unstructured data in the database.
So the performance stands different in Q2.

After pre-extraction and caching of the semantic information, we re-evaluate the overall performance, and the results are presented in Figure \ref{fig:expr_overview-cached}.
We found that PandaDB performs 100x to 1000x faster than the native implementations without any optimization.
As introduced before, extracting semantic information takes most of the time.
In the native solutions, data flow from one component to another costs much, especially when the data is large (unstructured data is also larger than structured data).
However, PandaDB executes these queries in an optimization data pipeline based on the optimized data storage. This improves the native solution greatly in the experiment and production environment. 

\subsection{Unstructured Data Storage Performance Evaluation}
\begin{figure*}
    \centering
    \scalebox{0.85}{\includegraphics[width=2\columnwidth]{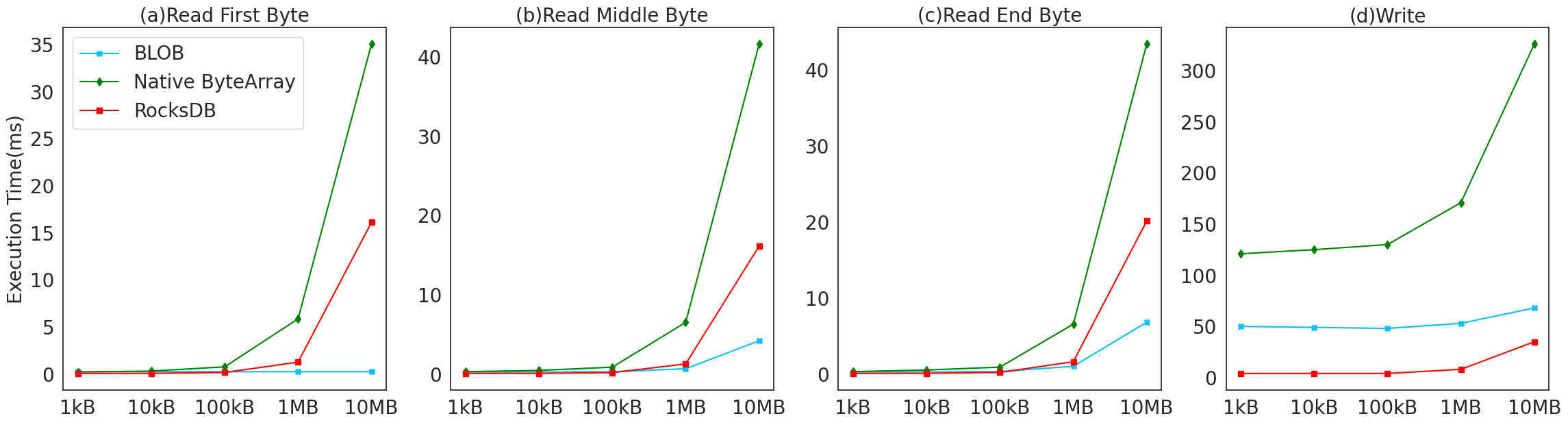}}
    \vspace{-1 em}
    \caption{Unstructured data storage performance evaluation}
    \vspace{-1 em}
    \label{fig:unstructured-data-storage}
\end{figure*}
To compare the reading and writing efficiency of the different storage formats of unstructured data, we conducted a read-write test on unstructured data. 
The data size is varied from 1KB to 10MB.
Considering the streaming data reading requirements, we record the time measured by the overhead to read the first byte, middle byte and end byte of unstructured data, specifically.
The results are shown in Figure \ref{fig:unstructured-data-storage}, BLOB by PandaDB performs over 10x faster than others. In addition, both the Neo4j and RocksDB share a similar performance in loading the first bytes of unstructured data since they need to load the whole unstructured item from the disk at first.

\subsection{Effect of Indexing}
In this section, we evaluate how the index improves the vector searching for the semantic information data. We build the index and evaluate the effectiveness of PandaDB while processing semantic information on different scales.
For instance, the vector dataset can be SIFT-1M~\cite{sift1M}, and SIFT-100M (1/10 of the SIFT1B~\cite{sift1B}).
In this experiment, we build the index for the input dataset, then execute kNN query to compute the recall and related query processing time.
It takes 39 seconds to build the index for SIFT-1M and 3556 seconds for SIFT-100M. 
Each kNN query is repeated 500 times, then the min, max and average values are recorded.
\begin{figure}
    \centering
    \scalebox{0.88}{\includegraphics[width=\linewidth]{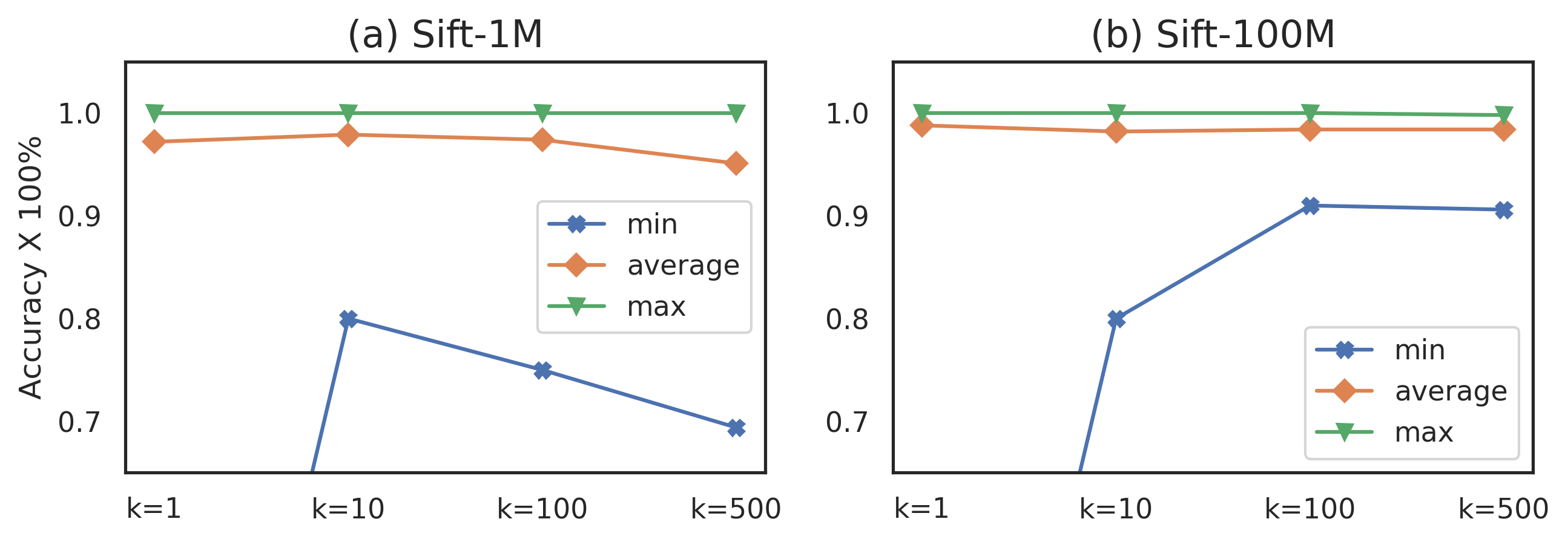}}
    \vspace{-1 em}
    \caption{Index recall evaluation on kNN search.}
    \vspace{-1 em}
    \label{fig:ex3-index-recall}
\end{figure}

\begin{figure}
    \centering
    \scalebox{0.88}{\includegraphics[width=\linewidth]{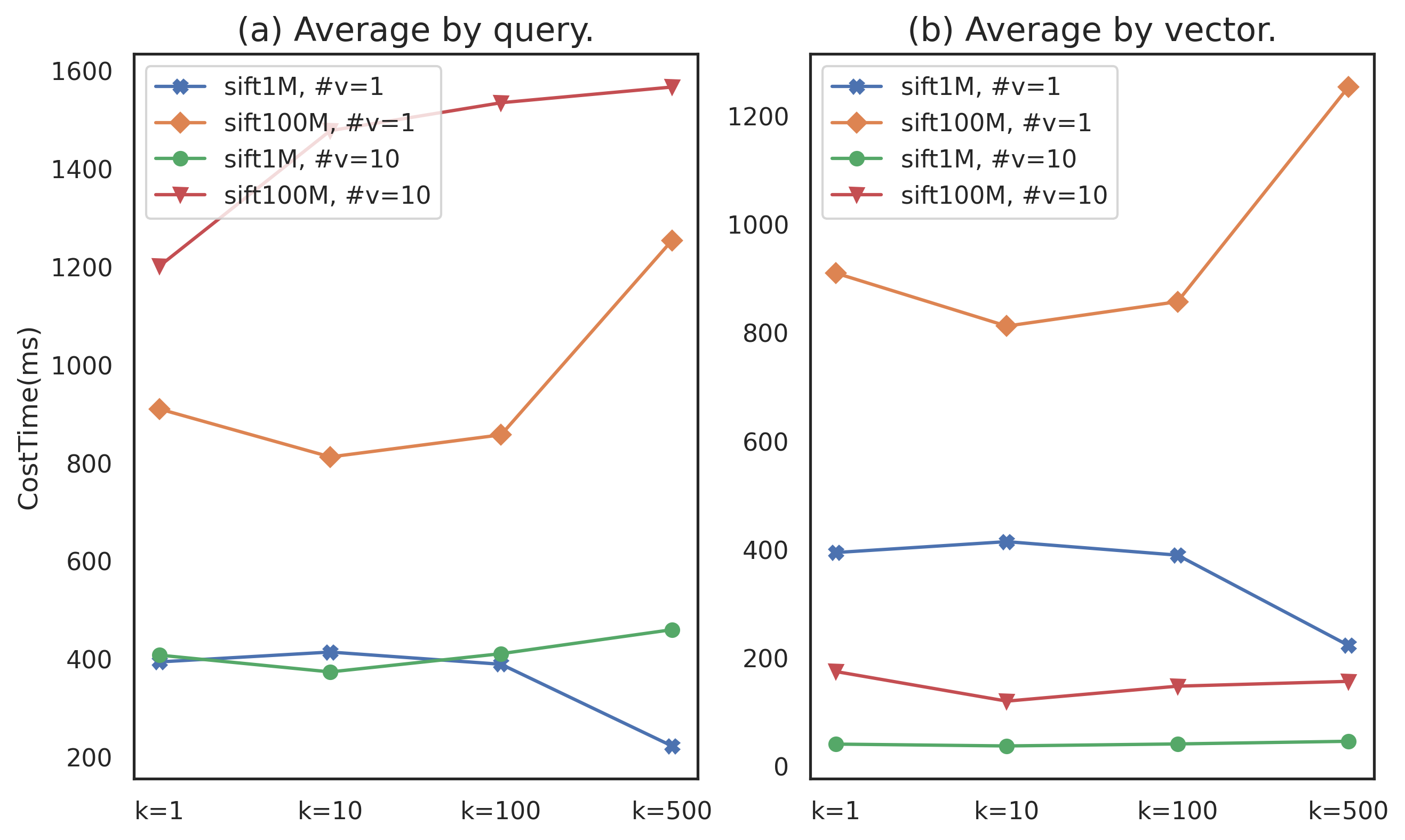}}
    \vspace{-1 em}
    \caption{Index performance evaluation on kNN search.}
    \vspace{-1.5 em}
    \label{fig:index-performance}
\end{figure}
Figure \ref{fig:ex3-index-recall} gives the recall of the built index. The average recall is higher than 0.95. This proves that the accuracy of the built index is acceptable in real applications.
Note that when k is set to one, the min recall rate is 0 because once an error occurs, the recall rate of the query will be 0. However, this is rare in real applications because the average rate is more than 0.95 in general.

Figure~\ref{fig:index-performance} shows the effect of the built index on query processing.
The ${\#v}$ stands for the count of vectors in a search. When a query includes multiple vectors, the average query time per vector is less than 0.2 second, which would meet the general demand of applications.
For comparison, we load all vectors into memory and perform the same queries using the traversal method. Each query on sift1M/sift100M takes about 5/500 seconds on average.
More details about the evaluation results are presented in the appendix of the tech report~\cite{tech_report_pandadb}. 

\section{Conclusion}
\label{sec:conclusion}
In this work, we extend the query language (namely \textit{CypherPlus}) to support query unstructured data in the graph database. We then provide an unified interface to meet requirements of semantic information extraction in the database system. 
In addition, we propose an optimization algorithm that is aware to unstructured data operation, and it optimize the query plan according to the meta information of semantic information (including whether it is cached/indexed or not). 
We also extend Neo4j's type system and develop a general \textit{BLOBValueManager} to manage unstructured data storage. 
Finally, we provide index to speedup the semantic information querying. The built system is widely used in different application related to manage unstructured and structured in a graph database system, and the experiment results show the built system can on average win 100x speedup against the system without optimization. 
\section{Appendix}
\subsection{Definition of Property Graph Model}
\begin{table}[ht!]
  \begin{center}
    \vspace{-1em}
    \caption{Summary of notational conventions}
    \vspace{-1em}
    \begin{tabular}{|l|c|c|}
    \hline
    \label{table-notation}
      \textbf{Concept} & \textbf{Notation} & \textbf{Set notation}\\
      \hline
      Property keys & $k$ & $\mathcal{K}$\\
      Sub-property keys & $sk$ & $\mathcal{SK}$ \\
      Relationship identifiers & $r$ & $\mathcal{R}$\ \\
      Node labels & $l$ & $\mathcal{L}$ \\
      Relationship types & $t$ & $\mathcal{T}$ \\
      Property values & $v$ & $\mathcal{V}$\\
     \tabincell{l}{Sub-property values \\ (Semantic information)} & $sv$ & $\mathcal{SV}$\\
      Unstructured data item & $ud$ & $S_{ud}$\\
    Sub-property extraction function & $\phi$ & $S_{\phi}$ \\
    Semantic space & $Sem$ & \\
    \hline
    \end{tabular}
  \end{center}
  \vspace{-2em}
\end{table}

We give the formal specification of the property graph data model referring to \cite{angles2008survey}, the notations are shown in Table~\ref{table-notation}.
Let $\mathcal{L}$ and $\mathcal{T}$ be countable sets of node labels and relationship types.
A property graph is a tuple G = $<N, R, K, src, tgt, \iota, \lambda, \tau>$ where:
\begin{itemize}
    \item $N$ is a finite subset of $\mathcal{N}$, whose elements are referred to as the nodes of G.
    \item $R$ is a finite subset of $\mathcal{R}$, whose elements are referred to as the relationships of G.
    \item $K$ is a finite subset of $\mathcal{K}$, whose elements are referred to as the properties of N and R.
    \item $src$: R $\rightarrow$ N is a function that maps each relationship to its source node.
    \item $tgt$: R $\rightarrow$ N is a function that maps each relationship to its target node.
    \item $\iota$: (N $\cup$ R) $\times$ $\mathcal{K}$ $\rightarrow$ $\mathcal{V}$ is a finite partial function that maps an identifier and a property key to a value.
    \item $\lambda$: N $\rightarrow$ $\mathcal{L}$ is a function that maps each node id to a finite set of labels.
    \item $\tau$: R $\rightarrow$ $\mathcal{T}$ is a function that maps each relationship identifier to a relationship type.
\end{itemize}

Following this definition, the graph in Figure~\ref{fig:example_graph} could be represented as G = $<N, R, K, src, tgt, \lambda, \iota, \tau>$:
\begin{itemize}
    \item $N$ = \{$n_{1}$,..., $n_{8}$\};
    \item $R$ = \{$r_{1}$,..., $r_{8}$\};
    \item $$ src = 
    \left\{
            \begin{array}{ccc}
              r_1 \mapsto n_1 & r_2 \mapsto n_1 & r_3 \mapsto n_1 \\
              r_4 \mapsto n_4 & r_5 \mapsto n_6 & r_6 \mapsto n_5 \\
              r_7 \mapsto n_2 & r_8 \mapsto n_1
            \end{array}
            \right\};
    $$
    \item $$ tgt = 
    \left\{
            \begin{array}{ccc}
            r_1 \mapsto n_2 & r_2 \mapsto n_3 & r_3 \mapsto n_4 \\
            r_4 \mapsto n_2 & r_5 \mapsto n_5 & r_6 \mapsto n_7 \\
            r_7 \mapsto n_7 & r_8 \mapsto n_8
            \end{array}
    \right\};
            $$
    \item 
        $\lambda$($n_1$)=$\lambda$($n_4$)=$\lambda$($n_6$)=$\lambda$($n_8$)=\{Person\},\\
        $\lambda$($n_2$)=$\lambda$($n_5$) = \{Team\},\\
        $\lambda$($n_3$)=\{Pet\},
        $\lambda$($n_7$)=\{Organization\};
    \item 
    $\iota$($n_{1}$, name) = Michael\ Jordan, 
    ...,
    $\iota$($n_{4}$, photo) = $<$\$image$>$;
    \item 
        $$
            \tau(r) = \left\{ 
            \begin{array}{lll}
                workFor & for & r\in \{r_1, r_4\} \\     
                hasPet & for & r\in \{r_2\} \\
                teamMate & for & r\in \{r_3, r_8\} \\
                coachOf & for & r\in \{r_5\} \\
                belongTo & for & r\in \{r_6, r_7\}
            \end{array}
            \right.
        $$
\end{itemize}
\subsection{Example of Cypher}
The following query statements show how to create and query data via Cypher for Figure\ref{fig:example_graph}.
\begin{lstlisting}[language=cypherplus]
-- Q1: Create two nodes and a relationship.
CREATE (jordan:Person{name: 'Michael Jordan'})
CREATE (scott:Person{name: 'Scott Pippen'})
CREATE (jordan)-[:teamMate]->(scott);
-- Q2: Query John's friend's name.
MATCH (jordan)-[:teamMate]->(n)
WHERE jordan.name='Michael Jordan'
RETURN n.name;
\end{lstlisting}

Q1 creates two nodes and builds a relationship, then two nodes are labeled with \textit{Person}, with the name 'Michael Jordan' and 'Scott Pippen', respectively.
Q2 retrieves the \textit{teamMate} relationship starts from the node with the name 'Micheal Jordan' and gets the related nodes' name property.
\subsection{Proof about the optimization foundation.}
The traditional Cost-Based-Optimization (CBO) methods reorder the operators in a query plan to find the one plan with minimal cost. An important precondition is that the relative order between operators does not affect the query results. That is, a filter operator always outputs the correct results, namely TP (true positive) and TN (true negative). However, while applying the filter operator to unstructured data processing, the filter operator first evaluates the semantic information extracted by the AI model, obtains the filter result, and then pushes the results to the later operator. As we know, the AI model is not one hundred percent accurate. It stands a certain probability of output FP (false positive) and FN (false negative). Therefore, when a filter operator of unstructured data is executed at first, its FP and FN output will become the input of its next operator. 
We prove that the filtering operator reordering does not affect the query results in a graph database. This conclusion makes it possible to optimize the query plan using CBO methods. 
We come to analyze the correctness to reorder the different operator for unstructured data and structured filtering in the graph database as below, this is important to obtain a more efficient query plan. 
\\
\textit{\textbf{Lemma1.}} \textit{The filtering operator reordering does not affect the query results in graph database.} \\
\label{lemma1}
Given a filter operator $f$, its input is $x$, the probability of its output TP/TN is $p$, and the probability of FP/FN is $q$. Then the filter operator can be expressed as:
\begin{equation}
    \label{equation_1}
    f(x)=p \cdot t(x) + q \cdot e(x)
\end{equation}
Then for the two filter operators $f_1$ and $f_2$, they can be expressed as:
\begin{equation}
\label{equation_2}
    \begin{aligned}
    f_1(x)=p_1 \cdot t_1(x) + q_1 \cdot e_1(x)
    \\
    f_2(x)=p_2 \cdot t_2(x) + q_2 \cdot e_2(x)
    \end{aligned}
\end{equation}
Then, in a query plan, execute $f_2$ first and then $f_1$, and the result can be expressed as:
\begin{equation}
\label{equation_3}
\begin{aligned}
f_1(f_2(x)) = & f_1(p_2 \cdot t_1(x) + q_1 \cdot e_1(x))
         \\ = & p_1 \cdot t_1(p_2 \cdot t_1(x) + q_1 \cdot e_1(x)) 
         \\ & + q_1 \cdot e_1(p_2 \cdot t_1(x) + q_1 \cdot e_1(x))
         \\ = & p_1 \cdot p_2 \cdot t_1(t_2(x)) + p_1 \cdot q_2 \cdot t_1(e_2(x))
         \\ & + p_2 \cdot q_1 \cdot e_1(t_2(x)) + q_1 \cdot q_2 \cdot e_1(e_2(x))
\end{aligned}
\end{equation}
Similarly, execute $f_1$ first and then $f_2$ to obtain the result:
\begin{equation}
\label{equation_4}
\begin{aligned}
f_2(f_1(x)) = & p_1 \cdot p_2 \cdot t_2(t_1(x)) + p_2 \cdot q_1 \cdot t_2(e_1(x))
         \\ & + p_1 \cdot q_2 \cdot e_2(t_1(x)) + q_1 \cdot q_2 \cdot e_2(e_1(x))
\end{aligned}
\end{equation}

Let $\Delta$ be the difference between equation \ref{equation_3} and equation \ref{equation_4}. If the value of delta is equal to zero, then equation \ref{equation_3} and equation \ref{equation_4} are equivalent.
\begin{equation}
\begin{aligned}
\Delta = & f_1(f_2(x)) - f_2(f_1(x)) 
    \\ = & p_1 \cdot p_2 \left[ t_1(t_2(x)) - t_2(t_1(x)) \right]
    \\ & + p_1 \cdot q_2 \left[ t_1(e_2(x)) - e_2(t_1(x)) \right]
    \\ & + p_2 \cdot q_1 \left[ e_1(t_2(x)) - t_2(e_1(x)) \right]
    \\ & + q_1 \cdot q_2 \left[ e_1(e_2(x)) - e_2(e_1(x)) \right]
    \\ = & p_1 \cdot q_2 \left[ t_1(\neg t_2(x)) - \neg t_2(t_1(x))\right]
    \\ & + p_2 \cdot q_1 \left[ \neg t_1(t_2(x)) - t_2(\neg t_1(x))\right]
    \\ & + q_1 \cdot q_2 \left[ \neg t_1(\neg t_2(x)) - \neg t_2(\neg t_1(x))\right]
    \\ = & 0
\end{aligned}
\end{equation}
\subsection{Build Index for Semantic Data}
Algorithm \ref{algo:indexing} shows how PandaDB builds index for semantic space composed of vectors.
For high dimensional vectors, we divide the space into $m$ buckets.
Each bucket has a core vector, and vectors are assigned to this bucket based on the closet distance. Suppose a kNN search task where k=1, the system first calculates the distances of the vector to each core vector, then selects the corresponding bucket of the nearest core vector.
Next, execute a linear search in this bucket, and find the nearest vector.
For datasets with a larger scale, we also offer the implementation of HNSW\cite{hnsw} and IVF\_SQ8\cite{ivf_sq8}. 
These two index algorithms perform better on larger datasets of vectors, and HNSW even supports dynamic insert. The inverted vector search is an ANNS(Approximate Nearest Neighbour Search).

\begin{algorithm}
    \SetKwFunction{Extract}{ExtractSemInfo}
    \SetKwFunction{BuildSpace}{GetSemSpace}
    \SetKwFunction{Pick}{PickBucket}
    \SetKwFunction{Batch}{BatchIndexing}
    \SetKwFunction{Dynamic}{DynamicIndexing}
    \SetKwProg{Fn}{Function}{:}{}
    \caption{Semantic Information Indexing Algorithm}
    \label{algo:indexing}
    \KwIn{Semantic Space $S$}
    \KwOut{Indexed Semantic Space}
    \Fn{\Pick{vec, B}}{
        D $\gets$ $\emptyset$\\
        \ForEach{bucket $\in$ B}{
            d $\gets$ distance(vec, bucket.core)\\
            D.insert(d, bucket)
        }
        bucket $\gets$ minByDis(D)\\
        \textbf{return} bucket
    }
    \Fn{\Batch{$S$}}{
        \If{S is $\emptyset$}{S $\gets$ $\BuildSpace{D, Schema, subPty}$}
        m $\gets$ count(S) \\
        B $\gets$ $\emptyset$ \Comment{Bucket Set}\\
        \While{size(B) $<$ m/100000}{ \Comment{100000 is an empirical value}\\
            bucket.core $\gets$ randomSelect(S) \\
            S.remove(bucket.core)
        }
        \ForEach{vec $\in$ Space}{
            bucket $\gets$ $\Pick(vec, B)$ \\
            bucket.insert(vec)
        }
        \textbf{return} B
    }
    \Fn{\Dynamic{d}}{
        i $\gets$ $\Extract{d, subPty, Schema}$ \\
        Space.insert(i) \\
        bucket $\gets$ $\Pick(i, B)$ \\
        bucket.insert(i) \\
        \textbf{return} B
    }
\end{algorithm}

\subsection{Native Solution Implementation}
\begin{figure*}
    \centering
    \includegraphics[width=\linewidth]{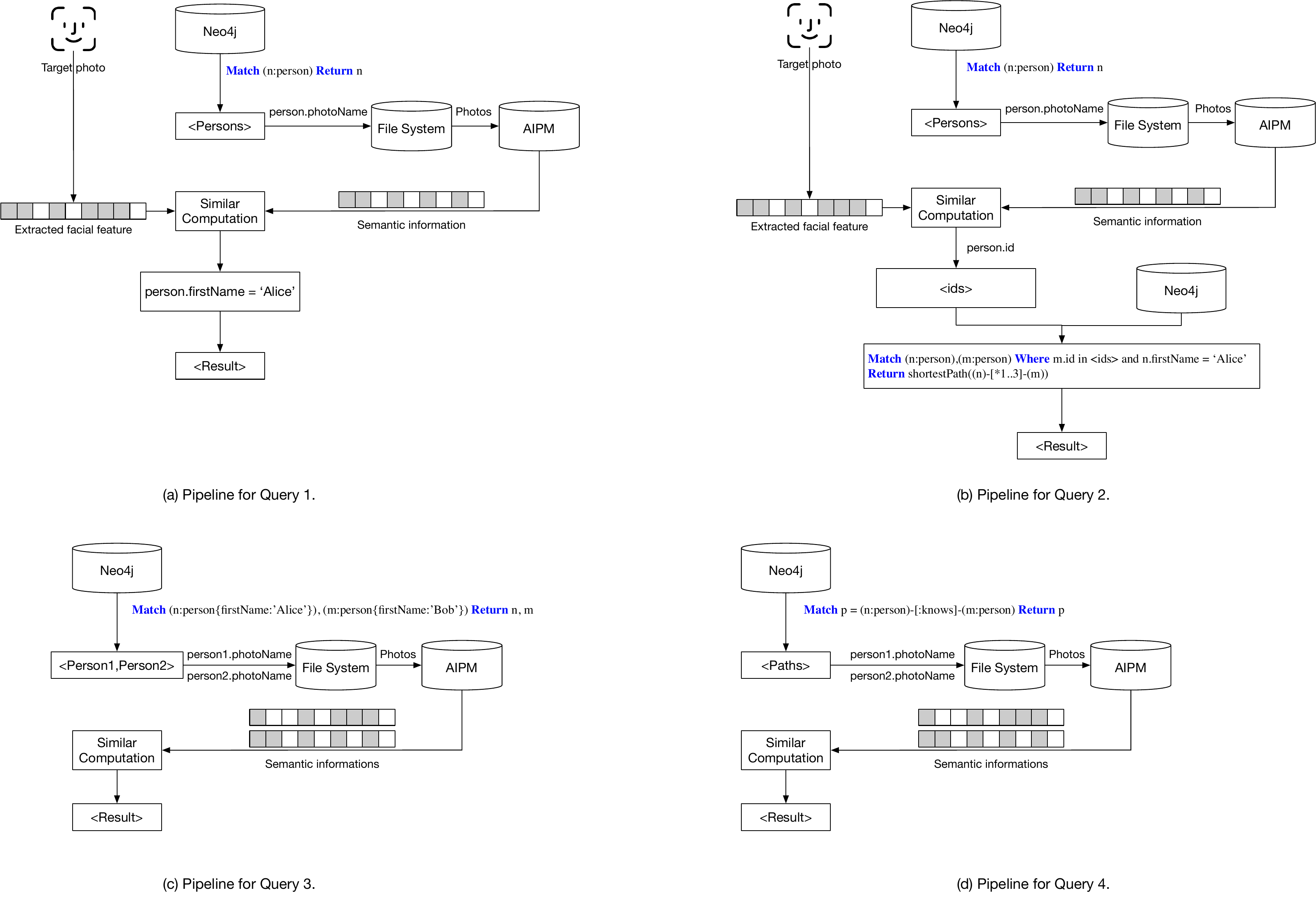}
    \caption{Native solution pipelines for the example queries.}
    \label{fig:native-solutions}
\end{figure*}

In this section, we give the details of native solution implementation. 
Figure~\ref{fig:native-solutions} gives the diagrams of the pipelines for the four example queries in section~\ref{subsubsec:query}, respectively.
\begin{itemize}[ht!]
    \item Q1: First extract the facial feature from the target photo. Then get all the photo's paths and retrieve the corresponding photos from the file system. Following, the AIPM extracts all the facial features (a.k.a, semantic information) from the photos. Then the scripts compute the semantic information with the target photo's facial feature, keep the results whose similarity exceed the threshold (actually 0.8 in this experiment). Finally, filter the nodes by the \textit{firstName} property and return the result.
    \item Q2: Similar to pipeline for Q1, extract the facial features of target photo and semantic information (a.k.a facial features) of the corresponding nodes. Then compute the similarity and filter by the threshold. Then execute the shortest path query in Neo4j, with the filtered output as the condition. Finally return the queried result.
    \item Retrieve the nodes by \textit{firstName} property and get the corresponding photos from the file system. Then get the semantic information of these photos from AIPM, and compute the similarity.
    \item Get all the paths meeting the query conditions, and get the photos of the start and end nodes. Then extract the semantic information of these photos and compute the similarity.
\end{itemize}
Note that in the experiment where semantic information are cached, the AIPM pre-extract the semantic information and use the local cache, instead of extracting the semantic information in real time.

\subsection{Optimization Comparison}
The results are shown in Figure\ref{fig:expr_overview-nocache} and Figure\ref{fig:expr_overview-cached}.
The features differ from one query to another, so the optimization efficiency differs.
There are two filters in Q1, one for structured data(filter by name), and the other for semantic information(filter by face feature).
The input of the first filter is all the property data in the database,
while the input for the second filter is the output of the first one.
Obviously, executing the filter for the name would make the semantic information filter extract fewer data than executing the name filter later.
While in Q2 and Q3, the number of semantic information to be extracted could not be narrowed down, so the optimization does not perform well.

When the semantic information is pre-extracted and cached, the optimization performs better in Q2.
In this case, the semantic information filter is slower than the structured property filter, so putting semantic information filtering behind can reduce the overhead.
In the case without cache, there is also this optimization logic.
However, when there is no cache, the extraction of semantic information takes much time, so the effect of this optimization is not apparent.

\subsection{Index Performance Evaluation}
We brought kNN search on the datasets(respectively, with k=1, 10, 100, 500).
For each k value, the experiment is repeated 500 times, recording the max, min, and average of the query accuracy.
The result is shown in Figure\ref{fig:ex3-index-recall}.
The average accuracy is stable above 0.95. When the K value is small, there are very few cases of low accuracy.

In order to evaluate the query speed of the index, we carried out experiments from the perspectives of single vector retrieval and batch vector retrieval.
For single vector retrieval, KNN retrieval is performed on one vector at a time, and the query time is recorded.
For batch vector retrieval, ten vectors are searched by KNN each time, and the query time is recorded.
Among them, the value of K is 1, 10, 100, and 500, respectively. For each k value, repeat 500 times and record the average value.
The results are shown in Figure \ref{fig:index-performance}, where the \#v means the number of vectors included in a query.
Figure \ref{fig:index-performance}(a) records the average time spent per query in 500 repeated experiments under different conditions.
Figure \ref{fig:index-performance}(b) records the average time spent per vector in a query, for queries with \#v = 1. The average time of each vector is the time of the query. For \#v = 10 queries, the average time per vector is 1 / 10 of the query time.
On the same dataset, the total time consumption of single query and batch query is very close and does not change significantly with the change of K value.
The average time consumption does not increase significantly with the increase of K value, which also enlightens us that we can reduce the average time consumption of each vector query by submitting batch query tasks.

\subsection{Cases Studies}
\subsubsection{\textbf{Academic graph disambiguation and mining}}
\begin{figure}
    \centering
    \includegraphics[width=\linewidth]{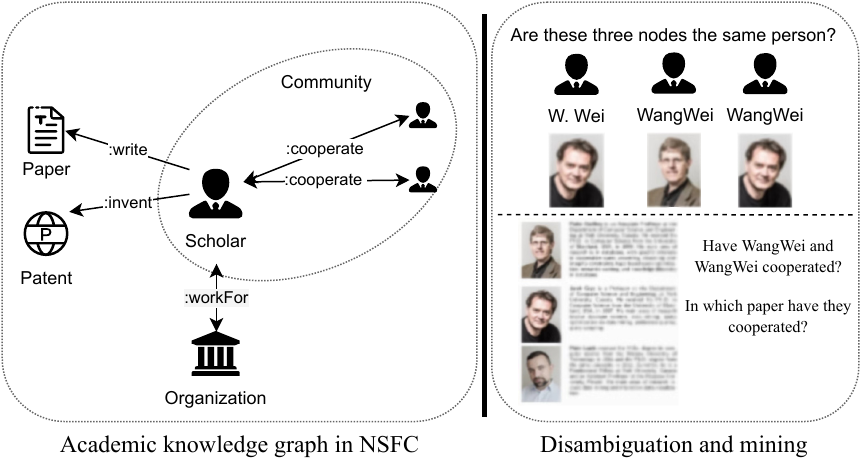}
    \caption{Academic graph disambiguation}
    \vspace{-1.5 em}
    \label{fig:use_case1}
\end{figure}

NSFC\\(National Natural Science Foundation of China) stores and manages data about scholars, published papers, academic affiliations and scientific research funds details.
Figure~\ref{fig:use_case1} shows the data overview in NSFC. There are about 1.5TB of data, with 2 million scholars.
Three example queries are shown in Figure\ref{fig:use_case1}. All of them involve unstructured semantic information. About sixty different types of queries similar to these three are carried on the system.
We use OCR technology to extract the author and scientific research organization information from the PDF files of the papers, then construct the corresponding association relationships between authors and their corresponding universities. This affiliation is used to build the connection between two graph nodes. Then we apply the GNN model based on the graph query operators over unstructured data~\cite{qiaoziyue}, and the average model training and prediction time are reduced by more than forty and twenty percent respectively. 

\subsubsection{\textbf{DoubanMovie system}}
When watching TV programs, viewers often look at an actor and cannot remember his name or what programs the actor has played.
PandaDB is deployed to help users to find the Superstar in \textit{DoubanMovie}\footnote{https://movie.douban.com/}, the biggest movie comment and review website in China. \textit{DoubanMovie} contains more than 100 million movies, superstars, comments and users.
We built a graph containing actors, movies, and participation relationships.
When the user submits a photo, PandaDB can find the superstars sharing a similar photo as the facial information of the input photo, then find the film in which the actor has played from the graph. This system is deployed and used in the production environment, and one demo video is in the link\footnote{https://github.com/Airzihao/Airzihao.github.io/blob/master/gif/demo.gif}

\bibliographystyle{IEEEtran}
\bibliography{reference}

\end{document}